\documentclass[envcountsame,oribibl]{./llncs}

\usepackage[latin1]{inputenc}
\usepackage{array}
\usepackage{amsmath}
\usepackage{amssymb}
\usepackage{latexsym}
\usepackage[all]{xy}
\usepackage{url}

\newcommand{\comment}[1]{}

\newcommand{\vide}{\emptyset}
\newcommand{\ie}{{\em i.e.} }
\newcommand{\eg}{{\em e.g.} }

\newcommand{\xt}{\{x\to t\}}

\newcommand{\FV}{\mr{FV}}

\renewcommand{\a}{\rightarrow}
\newcommand{\al}{\leftarrow}
\newcommand{\A}{\Rightarrow}
\newcommand{\alr}{\leftrightarrow}
\newcommand{\Alr}{\Leftrightarrow}

\newcommand{\ad}{\downarrow}

\renewcommand{\to}{\mapsto}

\renewcommand{\|}{\: | \:}
\newcommand{\tq}{\: ; \:}

\newcommand{\ex}{\exists}
\newcommand{\all}{\forall}

\newcommand{\st}{\star}
\newcommand{\sle}{\subseteq}

\newcommand{\sgt}{\supset}

\newcommand{\tlt}{\lhd}
\newcommand{\tgt}{\rhd}

\renewcommand{\b}{\beta}

\newcommand{\vep}{\varepsilon}

\renewcommand{\t}{\theta}

\newcommand{\la}{\lambda}

\newcommand{\s}{\sigma}
\newcommand{\Si}{\Sigma}

\newcommand{\mc}{\mathcal}

\newcommand{\mr}{\mathrm}

\newcommand{\cA}{\mc{A}}
\newcommand{\cB}{\mc{B}}

\newcommand{\cF}{\mc{F}}

\newcommand{\cN}{\mc{N}}
\newcommand{\cO}{\mc{O}}

\newcommand{\cR}{\mc{R}}
\newcommand{\cS}{\mc{S}}
\newcommand{\cT}{\mc{T}}
\newcommand{\cU}{\mc{U}}

\newcommand{\cW}{\mc{W}}
\newcommand{\cX}{\mc{X}}

\newcommand{\va}{{\vec{a}}}
\newcommand{\vb}{{\vec{b}}}
\newcommand{\vc}{{\vec{c}}}
\newcommand{\vd}{{\vec{d}}}

\newcommand{\vl}{{\vec{l}}}

\newcommand{\vt}{{\vec{t}}}
\newcommand{\vu}{{\vec{u}}}
\newcommand{\vv}{{\vec{v}}}

\newcommand{\vx}{{\vec{x}}}

\newcommand{\bnf}{\b n \! f}

\newcommand{\cqfd}{\hfill$\square$} % \blacksquare
\newenvironment{prf}{{\em Proof Sketch.}}{\cqfd} %\qed

\begin{document}

\title{On the confluence of $\la$-calculus\\ with conditional rewriting}

\author{Fr\'ed\'eric Blanqui\inst{1} 
\and Claude Kirchner\inst{1}
\and Colin Riba\inst{2}
}

% \authorrunning{Fr\'ed\'eric Blanqui\and Claude Kirchner\and Colin Riba}

\institute{INRIA \& LORIA\thanks{UMR 7503 CNRS-INPL-INRIA-Nancy2-UHP,
	Campus Scientifique, BP 239,
	54506 Vandoeuvre-l\`es-Nancy Cedex, France}
\and 
INPL \& LORIA}

\maketitle

\begin{abstract}
The confluence of untyped $\la$-calculus with {\em unconditional}
re\-writing has already been studied in various directions.  In this
paper, we investigate the confluence of $\la$-calculus with {\em
conditional} rewriting and provide general results in two directions.
First, when conditional rules are algebraic. This extends results of
M\"uller and Dougherty for unconditional rewriting. Two cases are
considered, whether beta-reduction is allowed or not in the evaluation
of conditions. Moreover, Dougherty's result is improved from the
assumption of strongly normalizing $\b$-reduction to weakly
normalizing $\b$-reduction. We also provide examples showing that
outside these conditions, modularity of confluence is difficult to
achieve. Second, we go beyond the algebraic framework and get new
confluence results using a restricted notion of orthogonality that
takes advantage of the conditional part of rewrite rules.
\end{abstract}

%%%%%%%%%%%%%%%%%%%%%%%%%%%%%%%%%%%%%%%%%%%%%%%%%%%%%%%%%%%%%%%%%%%%%%%%%%%
% introduction
%%%%%%%%%%%%%%%%%%%%%%%%%%%%%%%%%%%%%%%%%%%%%%%%%%%%%%%%%%%%%%%%%%%%%%%%%%%

\section{Introduction}

Rewriting \cite{dershowitz90book} and $\la$-calculus
\cite{barendregt84book} are two universal computation models which are
both used, with their own advantages, in programming language design
and implementation, as well as for the foundation of logical
frameworks and proof assistants. Among other things,
$\la$-calculus allows to manipulate abstractions and higher-order variables,
while rewriting is traditionally well suited for
defining functions over data-types and for dealing with equality.

Starting from Klop's work on higher-order rewriting and because of
their complementarity, many frameworks have been designed with a view
to integrate these two formalisms. This integration
has been handled
either by enriching first-order rewriting with higher-order
capabilities, by adding to $\la$-calculus algebraic features or, more
recently, by a uniform integration of both paradigms. In the first
case, we find the works on combinatory reduction systems
\cite{KlopOostromRaamsdonk} and other higher-order rewriting systems
\cite{nipkow91lics} each of them subsumed by van Oostrom 
and van Raamsdonk's axiomatization of HORS \cite{OR94}.
The second case concerns the more atomic
combination of $\la$-calculus with term rewriting
\cite{jouannaud91lics,blanqui05mscs} and the last category
the rewriting calculus \cite{rhoCalIGLP-I+II-2001,barthe03a}.

Despite this strong interest in the combination of both concepts, few works have
considered {\em conditional} higher-order rewriting in $\la$-calculus. This is
of particular interest for both computation and deduction. Indeed, conditional rewriting
appears to be very convenient when programming with rewrite rules and its combination
with higher-order features provides a quite agile background for the combination of
algebraic and functional programming. This is also of main use in proof assistants based
on the de Bruijn-Curry-Howard isomorphism where, as emphasized in {\em deduction modulo}
\cite{TPM-DHK-JAR-03,blanqui05mscs}, rewriting capabilities for defining functions and
proving equalities automatically is clearly of great interest when making large
proof developments.
Furthermore, while many confluence proofs
often rely on termination and local confluence, in some cases,
confluence may be necessary for proving termination (\eg with
type-level rewriting or strong elimination \cite{blanqui05mscs}). It
is therefore of crucial interest to have also criteria for the
preservation of confluence when combining conditional rewriting and
$\b$-reduction without assuming the termination of the combined
relation. In particular, assuming the termination of just one of the
two relations is already of interest.

The earliest work on preservation of confluence when combining
typed $\la$-calculus and first-order rewriting concerns the simple type
discipline \cite{breazu88lics} and the result has been extended to
polymorphic $\la$-calculus in \cite{breazu94ic}. Concerning untyped
$\la$-calculus, the result was shown in \cite{muller92ipl} for
left-linear rewriting.
It is extended as a modularity result for higher order rewriting in \cite{OR94}.
In \cite{dougherty92ic}, it is shown that 
left-linearity is not necessary 
provided that terms considered are strongly
$\b$-normalizable and are well-formed with respect to the declared
arity of symbols, a property that we call here {\em arity-compliance}.
Higher-order conditional rewriting is studied in
\cite{avenhaus94HOcondRew} and the confluence result relies on
joinability of critical pairs, hence on termination of the combined
rewrite relation.
Another form of higher-order conditional rewriting is considered in
\cite{taka93tlca}.
It concerns confluence results for a very general form of orthogonal systems.
These systems are related to those presented in Sect. \ref{sec-exclus}.
If modularity properties have been investigated
in the pure first-order conditional case (\eg \cite{Midd:JSC94,gramlich-tcs96}), 
to the best of our
knowledge, there was up to now no result on {\em preservation} of confluence
when $\b$-reduction is added to {\em conditional} rewriting.

In this paper, we study the confluence property of the combination of
$\b$-reduction with a confluent conditional rewrite system. This of
course should rely on a clear understanding of the conditional rewrite
relation under use and, as usual, the ways the matching is performed
and instantiated conditions are decided are crucial.

So, we start from $\la$-terms with curried constants and among them we
distinguish {\em applicative} terms that contain no abstraction and
{\em algebraic} terms that furthermore have no active variables, \ie
variables occurring in the left-hand side of an application. In this
paper, we always consider algebraic left-hand sides. So, rewriting
does not use higher-order pattern-matching but just syntactic
matching. Furthermore, we consider two rewrite relations induced by a
set of conditional rules. $\a_\cA$ is the conditional rewrite relation
where the conditions are checked {\em without} considering
$\b$-reduction and $\a_\cB$ is the conditional rewrite relation where
$\b$-reduction is allowed when evaluating the conditions. Then, we
study the confluence of the relations $\a_{\b\cup\cA}$ and
$\a_{\b\cup\cB}$, the respective combinations of $\a_{\cA}$ and
$\a_{\cB}$ with $\b$-reduction. This is made precise in
Sect.~\ref{sec-gendef} and accompanied of relevant examples.

We know that adding $\b$-reduction to a confluent non left-linear algebraic rewriting
system results in a non confluent relation. Of course, with conditional rewriting,
non-linearity can be simulated by linear systems. Extending the result of M\"uller
\cite{muller92ipl}, we prove in Sect.~\ref{sec-Aconfl} that confluence of
$\a_{\b\cup\cA}$ follows from confluence of $\a_\cA$ when conditional rules are
applicative, left-linear and do not allow their condition to test for equality of open
terms. Such rules are called {\em semi-closed}. We also adapt to conditional rewriting the
method of Dougherty \cite{dougherty92ic} and extend it to show that for a large set
of {\em weakly} $\b$-normalizing
terms, the left-linearity and semi-closed hypotheses can be dropped provided the
rules are algebraic and terms are arity-compliant.

We then turn in Sect.~\ref{sec-Bconfl} to the confluence modularity of $\a_{\b\cup\cB}$
for rules with algebraic right-hand side. In this case, we show that arity-compliance is a
sufficient condition to deduce confluence of $\a_{\b\cup\cB}$ from confluence of
$\a_{\b\cup\cA}$ (hence of $\a_\cA$). This is done first for left-linear semi-closed
systems, a restriction that we also show to be superfluous when considering only 
{\em weakly} $\b$-normalizing terms.

The case of non-algebraic rules is handled in Sect.~\ref{sec-exclus}. Such rules can
contain active variables and abstractions in right-hand sides or in conditions (but still
not in left-hand sides). In this case, the confluence of $\a_{\b\cup\cB}$ no more follows
from the confluence of $\a_\cA$ nor of $\a_{\b\cup\cA}$.
We show that the confluence
of $\a_{\b\cup\cB}$ holds under a syntactic condition, called {\em
orthonormality} ensuring that if two rules overlap at a non-variable position, then their
conditions cannot be both satisfied.
An orthonormal system is therefore orthogonal, and the confluence of $\a_{\cB \cup \b}$
follows using usual proof methods.

\medskip

We assume some familiarity with $\la$-calculus \cite{barendregt84book} and conditional
rewriting \cite{dershowitz90tcs,ohlebusch02book} but we recall the main notations in the
next section.  By lack of place, the main proofs are only sketched here. They are detailed
in~\cite{blanqui05b}.

\comment{
We assume some familiarity with $\la$-calculus \cite{barendregt84book} and
conditional rewriting \cite{dershowitz90book,dershowitz90tcs,ohlebusch02book} but we
recall the main notations in the next section.
%% For referee usage, the majority of proofs are detailed in Appendices.
A version of this paper with full proofs is
available at \url{http://www.loria.fr/~riba}.
}

%%%%%%%%%%%%%%%%%%%%%%%%%%%%%%%%%%%%%%%%%%%%%%%%%%%%%%%%%%%%%%%%%%%%%%%%%%%
% general definitions
%%%%%%%%%%%%%%%%%%%%%%%%%%%%%%%%%%%%%%%%%%%%%%%%%%%%%%%%%%%%%%%%%%%%%%%%%%%

\section{General definitions}
\label{sec-gendef}

This section introduces the main notions of the paper.
We use $\la$-terms with curried constants.

%%%%%%%%%%%%%%%%%%%%%%%%%%%%%%%%%%%%%%%%%%%%%%%%%%%%%%%%%%%%%%%%%%%%%%%%%%%
% terms
%%%%%%%%%%%%%%%%%%%%%%%%%%%%%%%%%%%%%%%%%%%%%%%%%%%%%%%%%%%%%%%%%%%%%%%%%%%

\begin{definition}[Terms]
We assume given a set $\cF$ of {\em function symbols} and an infinite
set $\cX$ of variables. The set $\cT$ of {\em terms} is inductively
defined as follows:
\[
\begin{array}{r c l}
t,u \in \cT & ::= & f \in \cF \| x \in \cX \| t u \| \la x. t
\end{array}
\]
A term is {\em applicative} if it contains no abstraction and {\em
algebraic} (``not variable-applying'' in \cite{muller92ipl}) if it
furthermore contains no subterm of the form $x t$ with $x\in\cX$. We
use $\vt$ to denote a sequence of terms $t_1,\ldots,t_n$ of length
$|\vt|=n$.
\end{definition}

As usual, terms are considered modulo $\alpha$-conversion. Let
$\FV(t)$ be the set of variables free in $t$. We denote by $t\s$ the
capture-avoiding application of the
substitution $\s$ to the term $t$. By $\{\vx\to\vt\}$,
we denote the substitution $\s$ such that $x_i\s=t_i$. As usual,
positions in a term are strings over $\{1,2\}$. The subterm of $t$ at
position $p$ is denoted by $t|_p$. If $t$ is applicative, the 
replacement of $t|_p$ by some term $u$ is
denoted by $t[u]_p$. A {\em context} is a term with exactly one free
occurrence of a distinguished variable $[]$. If $C$ is an applicative 
context then $C[t]$ stands for $C[t]_p$, where $p$ is the position of 
$[]$ in $C$.

A rewrite relation is a binary relation on terms $\a$ which is closed by
term formation rules :
if $s \a t$ then $\la x.s \a \la x.t$, $su \a tu$ and $us \a ut$ ; 
and by substitution : $s \a t$ implies $s\s \a t\s$.
Its inverse is denoted by $\al$; its reflexive 
closure by $\a^=$; its reflexive
and
transitive closure by $\a^*$; and its reflexive, symmetric and
transitive closure by $\alr^*$. The {\em joinability} relation is
${\ad}={\a^*\al^*}$. The $\b$-reduction relation is the smallest
rewrite relation $\a_\b$ such that $(\la x.s)t\a_\b s\xt$. A term $t$
$\a$-{\em rewrites} (or $\a$-{\em reduces}) to $u$ if $t\a^* u$
(we omit $\a$ when clear from the context).
We write $\a_{R\cup S}$ for the union of the
relations $\a_R$ and $\a_S$. We call {\em parallel rewrite relation}
any reflexive rewrite relation $\tgt$ 
closed by {\em parallel application} : $[s \tgt s' \,\&\, t \tgt t'] \A st \tgt s't'$.

% end comment

%%%%%%%%%%%%%%%%%%%%%%%%%%%%%%%%%%%%%%%%%%%%%%%%%%%%%%%%%%%%%%%%%%%%%%%%%%%
% conditional rewrite systems
%%%%%%%%%%%%%%%%%%%%%%%%%%%%%%%%%%%%%%%%%%%%%%%%%%%%%%%%%%%%%%%%%%%%%%%%%%%

We now introduce conditional rewriting. Let us emphasize that we
consider first-order syntactical matching.

\begin{definition}[Conditional rewriting]
\label{def-cond-rew}
A {\em conditional rewrite system} $\cR$ is a set of
conditional rewrite rules\footnote{The symbol $=$ does not
need to be interpreted by a symmetric relation.}:
\[
d_1= c_1 \land \dots \land d_n= c_n \sgt l\a r
\]
where $l$ is a non-variable algebraic term, $d_i$, $c_i$ and $r$ are
arbitrary terms and $\FV(d_i,c_i,r)\sle\FV(l)$. A system is {\em
right-applicative} (resp. {\em right-algebraic}) if all its right-hand
sides are applicative (resp. algebraic). A system is {\em applicative}
(resp. {\em algebraic}) if all its rules are made of applicative
(resp. algebraic) terms.

The {\em join} rewrite relation induced by $\cR$ is usually
defined as $\a_\cA= \bigcup_{i\geq
0}\a_{\cA_i}$ \cite{ohlebusch02book} where $\a_{\cA_0}=\vide$ and
for all $i \geq 0$, $\a_{\cA_{i+1}}$ is the smallest rewrite relation such that
for all rule $\vd = \vc \sgt l \a r \in \cR$, for all substitution $\s$,
if $\vd\s \ad_{\cA_i} \vc\s$ then $l\s \a_{\cA_{i+1}} r\s$.
This relation is sometimes called the {\em standard conditional rewrite
relation}.

We define the {\em $\b$-standard} rewrite relation induced by $\cR$ as
$\a_\cB= \bigcup_{i\geq 0}\a_{\cB_i}$ where $\a_{\cB_0}=\vide$ and
for all $i \geq 0$, $\a_{\cB_{i+1}}$ is the smallest rewrite relation such that
for all rule $\vd = \vc \sgt l \a r \in \cR$, for all $\s$,
if $\vd\s \ad_{\cB_i \cup \b} \vc\s$ then $l\s \a_{\cB_{i+1}} r\s$.

If $\a_{\cA_i}$ is confluent for all $i \geq 0$, we say that
$\a_{\cA}$ is {\em level confluent}. It is {\em shallow confluent}
when $\a^*_{\cA_i}$ and $\a^*_{\cA_j}$ commute for all $i,j \geq 0$.
\end{definition}

Other forms of conditional rewriting appear in the literature
\cite{dershowitz90tcs}. {\em Natural} rewriting is obtained by taking
$\alr_\cA^*$ instead of $\ad_\cA$ in the evaluation of
conditions. {\em Oriented} rewriting is obtained by taking
$\a_\cA^*$. A particular case of both standard and oriented rewriting
is {\em normal} rewriting, in which the terms $\vc$ are closed and in
$\a_\cA$-normal form.

%%%%%%%%%%%%%%%%%%%%%%%%%%%%%%%%%%%%%%%%%%%%%%%%%%%%%%%%%%%%%%%%%%%%%%%%%%%
% some examples 
%%%%%%%%%%%%%%%%%%%%%%%%%%%%%%%%%%%%%%%%%%%%%%%%%%%%%%%%%%%%%%%%%%%%%%%%%%%

%\section{Some examples}
\label{sec-ex}
\paragraph{\bf Examples.}
We begin by some basic functions on lists.
\[
%\begin{array}{c}
\begin{array}{c !{\qquad} c !{\qquad} c}
	\begin{array}{l !{\; \a \;} l}
	{\sf car}\; (x :: l) 	& x		\\
	{\sf car}\; [\;]		& {\sf err}
	\end{array}
&
	\begin{array}{l !{\; \a \;} l}
	{\sf cdr}\; (x :: l) 	& l		\\
	{\sf cdr}\; [\;]		& {\sf err}
	\end{array}
&
	\begin{array}{l !{\; \a \;} l}
	{\sf get}\; l\;  0 		& {\sf car}\; l				\\
	{\sf get}\; l\; ({\sf s}\; n)	& {\sf get}\; ({\sf cdr}\; l)\; n
	\end{array}
\end{array}
\]
\[
\begin{array}{c !{\qquad} c}
	\begin{array}{l !{\; \a \;} l}
	{\sf len}\; [\;]	& 0	\\
	{\sf len}\; (x :: l)& {\sf s}\; ({\sf len}\; l)
	\end{array}
&
	\begin{array}{l c l c l c l}
	      &     &          &       & {\sf filter}\; p \; [\;] & \a & [\;] \\
	p\; x & = & {\sf tt} & \sgt  & {\sf filter}\; p \; (x :: l)
		& \a & x :: ({\sf filter}\; p \; l) \\
	p\; x & = & {\sf ff} & \sgt & {\sf filter}\; p \; (x :: l)
		& \a & {\sf filter}\; p \; l
	\end{array}
\end{array}
%\end{array}
\]
Define $>$ with 
$>\, ({\sf s}\, x)\;  {\sf 0}      ~\a~ {\sf tt}$, 
$>\,  {\sf 0}\;       y            ~\a~ {\sf ff}$ and 
$>\, ({\sf s}\, x)\; ({\sf s}\, y) ~\a~ >\, x\; y$.
We can now define {\sf app} such that
${\sf app}\; f \; n\; l$ applies $f$ to the $n$th element of $l$. It uses {\sf ap}
as an auxiliary function:
\[
\begin{array}{c !{\quad} c}
\begin{array}{l !{\; = \;} l !{\; \sgt\; } l !{\; \a \;} l}
> ({\sf len}\; l)\; n & {\sf tt}  & {\sf app}\; f\; n\; l & 
	{\sf ap}\; f\; n\; l \\
> ({\sf len}\; l)\; n & {\sf ff}  & {\sf app}\; f\; n\; l &
	{\sf err} \\
\end{array}
&
\begin{array}{l !{\; \a \;} l}
{\sf ap}\; f\; 0            \; l & f\; ({\sf car}\; l) :: {\sf cdr}\; l \\
{\sf ap}\; f\; ({\sf s}\; n)\; l & {\sf car}\; l :: {\sf ap}\; f \; n\; ({\sf cdr}\; l)
\end{array}
\end{array}
\]
We represent
first-order terms as trees with nodes ${\sf nd}\; y\; l$ where
$y$ is intended to be a label and $l$ the list of sons.

Positions are
lists of integers and ${\sf occ}\; u\; t$ tests if $u$ is an occurrence
of $t$. We define it with ${\sf occ}\; [\;]\; t \;\a\; {\sf tt}$ and
\[
\begin{array}{l !{\; = \;} l !{\; \sgt \;} l !{\; \a \;} l}
>  ({\sf len}\; l)\; x & {\sf ff} &	 
	{\sf occ}\;  (x :: o)\; ({\sf nd}\; y\; l) & {\sf ff} \\
> ({\sf len}\; l)\; x  & {\sf tt}  &
	{\sf occ}\; (x :: o)\; ({\sf nd}\; y\; l) &
	{\sf occ}\; o\; ({\sf get}\; l\; x)
\end{array}
\]
To finish, ${\sf rep}\; t\; o\; s$
replaces by $s$ the subterm of $t$ at occurrence $o$.
Its rules are
${\sf occ}\; u \; t = {\sf tt} \sgt {\sf rep}\; t \; o \; s \a {\sf re}\; t \; o \; s$
and 
${\sf occ}\; u \; t = {\sf ff} \sgt {\sf rep}\; t \; o \; s \a {\sf err}$.
\comment{
\[
\begin{array}{c !{\qquad} c}
	\begin{array}{l !{\; = \;} l !{\; \sgt \;} l !{\; \a \;} l}
	{\sf occ}\; u \; t & {\sf tt} & {\sf rep}\; t \; o \; s & {\sf re}\; t \; o \; s \\ 
	\end{array}
&	
	\begin{array}{l !{\; = \;} l !{\; \sgt \;} l !{\; \a \;} l}
	{\sf occ}\; u \; t & {\sf ff} & {\sf rep}\; t \; o \; s & {\sf err} \\ 
	\end{array}
\end{array}
\]
The rules
${\sf re}\; ({\sf nd}\; y\; l) \; (x :: o)\; s~\a~
{\sf nd}\; y \; ({\sf app}\; 
		(\la z.{\sf re}\;z \; o\; s)\; x \; l)$
and 
${\sf re}\; s\; [\,] \; t ~\a~ s$ 
define the auxiliary function {\sf re}.
}
The rules
${\sf re}\; s\; [\,] \; t ~\a~ s$ 
and
${\sf re}\; ({\sf nd}\; y\; l) \; (x :: o)\; s~\a~
{\sf nd}\; y \; ({\sf app}\; 
		(\la z.{\sf re}\;z \; o\; s)\; x \; l)$
define the function {\sf re}.

The system {\sf Tree} that consists of rules defining {\sf car} {\sf
cdr}, {\sf get}, {\sf len} and {\sf occ} is algebraic.  Rules of {\sf
app} and {\sf ap} are right-applicatives and those for {\sf filter}
contain in their conditions the variable $p$ in active position.  {\em
This} definition of {\sf re} involves a $\la$-abstraction in a right
hand side. In Sect.~\ref{sec-exclus}, we prove confluence of the
relation $\a_{\b \cup \cB}$ induced by the whole system.

\comment{
\paragraph{\bf A remark on curryfication.}
Kahrs \cite{kahrs95jsc}, shows that curryfication of first order TRS preserve confluence.
The proof can be easily extended to
conditional rewriting.
However, {\em ground}
confluence is not preserved by curryfication:
the ground confluent TRS ${\rm g}(x) \a x$, ${\rm g}(x) \a a$ leads to
the unjoinable peak ${\sf g} \al {\sf g}\, {\sf g} \a {\sf a}$.
}

%%%%%%%%%%%%%%%%%%%%%%%%%%%%%%%%%%%%%%%%%%%%%%%%%%%%%%%%%%%%%%%%%%%%%%%%%%%
% confluence of beta union A
%%%%%%%%%%%%%%%%%%%%%%%%%%%%%%%%%%%%%%%%%%%%%%%%%%%%%%%%%%%%%%%%%%%%%%%%%%%

\section{Confluence of $\a_\b$ with conditional rewriting}
\label{sec-Aconfl}

In this section, we study the confluence of $\a_{\b\cup\cA}$. The
simplest result is the preservation of confluence when $\cR$ can not
check arbitrary equalities (Sect.~\ref{par-Alin}). In Sect.~\ref{par-Aweak}, 
we consider more general systems and prove that
the confluence of $\a_{\b\cup\cA}$ follows from the confluence of
$\a_\cA$ on terms having a $\b$-normal form of a peculiar kind.

In \cite{muller92ipl}, M\"uller shows that the union of $\b$-reduction
and the rewrite relation $\a_\cA$ induced by a left-linear
non-conditional applicative system is confluent as soon as $\a_\cA$ is.
This result is generalized as modularity result for higher-order
rewriting in \cite{OR94}.

The importance of left-linearity is known since Klop \cite{klop}.
We exemplify it with Breazu-Tannen's
counter-example \cite{breazu88lics}. The rules ${\sf -}\; x\; x ~\a~
{\sf 0}$ and ${\sf -}\; ({\sf s}\; x) \; x ~\a~ {\sf s}\; {\sf 0}$ are
optimization rules for minus. Together with usual rules defining this
function, they induce a confluent rewrite relation.  With the
fixpoint combinators of the $\la$-calculus, we can build a term $Y
\a^*_\b {\sf s}\;Y$. This term makes the application of the two rules
above possible on $\b$-reducts of $- \; Y \; Y$, leading to an
unjoinable peak~: $0 \al_\cA -\; Y \; Y \a_\b^* - \; ({\sf s}\; Y)
\; Y \a_\cA {\sf s}\; 0$.

With conditional rewriting, we do not need non-linear matching to
distinguish $-\; ({\sf s} \; x) \; x$ from $- \; x \; x$, since this
can be done within the conditions. The previous system can be encoded
into a left-linear conditional system with the rules $x \,=\, y
\,\sgt\, - \, x \, y \,\a\, 0 $ and ${\sf s}\, x \,=\, y \,\sgt\, -
\, x \, y \,\a\, {\sf s}\, {\sf 0}$. Of course, the relation $\a_\cA$
is still confluent. However, the same unjoinable peak starting from
$-\; Y\; Y$ makes fail the confluence of $\a_{\b\cup\cA}$. 

\comment{
Here the cause
is not the interaction of $\a_\b$ with first-order matching. Instead,
the problem comes from the ability of fixpoints to build terms that
change the algebraic meaning of rewrite rules. Indeed, non-confluence
of $\a_{\b\cup\cA}$ says here that in presence of full $\la$-calculus,
rewriting cannot safely express the distinction between $t$ and ${\sf
s}\; t$.
}

There are two ways to overcome the problem: limiting the power of
rewriting or limiting the power of $\b$-reduction. The first way is
treated in Sect.~\ref{par-Alin}, in which we limit the comparison power
of conditional rewriting by restricting ourselves to left-linear and {\em semi-closed}
systems. This can also be seen as a way, from the point of view of
rewriting, to isolate the effect of fixpoints: since two distinct
occurrences of $Y$ can not be compared, they can be unfolded
independently from each other.

Then, in Sect.~\ref{par-Aweak}, we limit the power of $\a_\b$ by
restricting ourselves to sets of terms having a special kind of
$\b$-normal-form. This amounts to only consider terms in which
fixpoints do not have the ability to modify the result of
$\a_{\b\cup\cA}$. In fact, it is sufficient that they do not modify
the result of $\a_\b$ alone. More precisely, fixpoints are allowed
when they are eliminated by head $\b$-reductions.

%%%%%%%%%%%%%%%%%%%%%%%%%%%%%%%%%%%%%%%%%%%%%%%%%%%%%%%%%%%%%%%%%%%%%%%%%%%
% left-linear systems
%%%%%%%%%%%%%%%%%%%%%%%%%%%%%%%%%%%%%%%%%%%%%%%%%%%%%%%%%%%%%%%%%%%%%%%%%%%

\subsection{Confluence of left-linear semi-closed systems}
\label{par-Alin}

We now introduce semi-closed systems.

\begin{definition}[Semi-closed systems]
A system is {\em semi-closed} if in every rule $\vd=\vc\sgt
l\a r$, each $c_i$ is algebraic and closed.
\end{definition}

%%%%%%%%%%%%%%%%%%%%%%%%%%%%%%%%%%%%%%%%%%%%%%%%%%%%%%%%%%%%%%%%%%%%%%%%%%%
% beta parallel
%%%%%%%%%%%%%%%%%%%%%%%%%%%%%%%%%%%%%%%%%%%%%%%%%%%%%%%%%%%%%%%%%%%%%%%%%%%

The system {\sf Tree} of Sect.~\ref{sec-ex} is left-linear and semi-closed.
Given a semi-closed left-linear system, we show that confluence of
$\a_{\b\cup\cA}$ follows from confluence of $\a_\cA$. This follows
from a weak commutation of $\a_\cA$ and Tait and Martin-L\"of
$\b$-parallel reduction relation $\tgt_\b$, defined as the smallest
parallel rewrite relation (Sect.~\ref{sec-gendef}) closed by the rule
$(beta)$ \cite{barendregt84book}:
\[
(beta) \quad \dfrac{s \tgt_\b s' \quad t \tgt_\b t'}{(\la x.s)t \tgt_\b s'\{x \to t'\}}
\]
We will use some well known properties of $\tgt_\b$. If
$\s \tgt_\b \s'$ then $s\s \tgt_\b s\s'$; this is the one-step
reduction of parallel redexes. We can also simulate $\b$-reduction:
$\a_\b \sle \tgt_\b \sle \a^*_\b$. And third, $\tgt_\b$ has the
diamond property: $\tlt_\b \tgt_\b \sle \tgt_\b \tlt_\b$.
This corresponds to the fact that any complete development of 
$\a_\b$ can be done in {\em one} $\tgt_\b$-step.

M{\"u}ller \cite{muller92ipl} uses a weaker parallelization of $\a_\b$:
its relation is defined w.r.t.\ the applicative structure of terms only and does not
reduces in one step nested $\b$-redexes. 
Consequently, it does not enjoy the diamond property
on which we rely in Sect.~\ref{sec-Bconfl}.
Nested parallelizations (corresponding to complete developments) are already
used in \cite{OR94} for their confluence proof of HORS.
However, our method inherits more from \cite{muller92ipl} than~\cite{OR94}, as
we use complete developments of $\a_\b$ only, whereas
complete developments of $\a_\b$ {\em and} of $\a_{\cA}$ are used 
for the modularity result of \cite{OR94}.

%%%%%%%%%%%%%%%%%%%%%%%%%%%%%%%%%%%%%%%%%%%%%%%%%%%%%%%%%%%%%%%%%%%%%%%%%%%
% commutation of \cA and beta parallel
%%%%%%%%%%%%%%%%%%%%%%%%%%%%%%%%%%%%%%%%%%%%%%%%%%%%%%%%%%%%%%%%%%%%%%%%%%%

\begin{proposition}
\label{prop-commut-rule}
Let $\cR$ be a semi-closed, left-linear and right-applicative system and
assume that $\a^*_{\cA_{i-1}}$ commutes with
$\a^*_\b$. For any rule $\vd = \vc \sgt l \a r \in \cR$ and
substitution $\s$, if $u \tlt_\b l\s \a_{\cA_i} r\s$, then there
exists $\s'$ such that $u = l\s' \a_{\cA_i} r\s' \tlt_\b r\s$.
\end{proposition}

\begin{prf}
Since $l$ is algebraic and linear, there is a substitution $\s'$ such
that $\s\tgt_\b\s'$ and $u=l\s'$. It follows that $r\s\tgt_\b r\s'$
and it remains to show that $\vd\s'\ad_{\cA_{i-1}} \vc\s'$. Since
$l\s\a_{\cA_i} r\s$, there is $\vv$ such that $\vd\s\a_{\cA_{i-1}}^*
\vv\al_{\cA_{i-1}}^* \vc\s$. Thus, $\vd\s\tgt_\b^* \vd\s'$ and, 
by assumption, there is
$\vv'$ such that $\vd\s'\a_{\cA_{i-1}}^* \vv'\tlt_\b^* \vv$. Since
$\vc$ is algebraic and closed, we have $\vc\s=\vc$ and $\vv$ in
$\b$-normal form. Hence, $\vv'=\vv$ and $\vd\s'\ad_{\cA_{i-1}}\vc$.
\end{prf}

\begin{lemma}[Commutation of $\a_\cA$ and $\tgt_\b$]
\label{lem-commut}
If $\cR$ is a semi-closed left-linear right-applicative system, then
${\tlt_\b^*\a_\cA^*} \sle {\a_\cA^*\tlt_\b^*}$.
\end{lemma}

\begin{prf}
The result follows from the commutation of $\a^*_{\cA_i}$ and
$\tgt^*_\b$ for all $i \geq 0$. The case $i=0$ is trivial. For $i>0$,
there are three steps.
First, we show by induction on the definition of the parallel rewrite relation
$\tgt_\b$ that if $u \tlt_\b s \a_{\cA_i} t$ then there exists $v$ such that
$u \a^*_{\cA_i} v \tlt_\b t$.
If $u$ is $s$ this is obvious. If $s$ is an abstraction, the result follows from 
induction hypothesis (IH) and the context
closure of $\a_{\cA_i}$ (CC).
If $s = s_1 s_2$, there are two cases: if
$t = t_1 t_2$ with $s_k \a^=_{\cA_i}  t_k$ then we conclude
by (IH) and (CC).
Otherwise, we use Prop. \ref{prop-commut-rule}.

Second, use induction on the number of
$\cA_i$-steps to show that ${\tlt_\b \a^*_{\cA_i}} \sle
{\a^*_{\cA_i} \tgt_\b}$. Finally, to conclude that ${\tlt_\b^*
\a_{\cA_i}^*} \sle {\a_{\cA_i}^*\tlt_\b^*}$, use an induction on the number
of $\tgt_\b$-steps.
\end{prf}

%%%%%%%%%%%%%%%%%%%%%%%%%%%%%%%%%%%%%%%%%%%%%%%%%%%%%%%%%%%%%%%%%%%%%%%%%%%
% confluence of beta union \cA
%%%%%%%%%%%%%%%%%%%%%%%%%%%%%%%%%%%%%%%%%%%%%%%%%%%%%%%%%%%%%%%%%%%%%%%%%%%

A direct application of Hindley-Rosen's Lemma offers then the preservation of confluence.

\begin{theorem}[Confluence of $\a_{\b\cup\cA}$]
\label{thm-conflA}
Let $\cR$ be a semi-closed left-linear right-applicative system.
If $\a_\cA$ is confluent then so is $\a_{\b\cup\cA}$.
\end{theorem}

For the system {\sf Tree} of Sect.~\ref{sec-ex}, the relation $\a_\cA$ is
confluent. As the rules are left-linear and semi-closed, Theorem
\ref{thm-conflA} applies and $\a_{\b\cup\cA}$ is confluent.

%%%%%%%%%%%%%%%%%%%%%%%%%%%%%%%%%%%%%%%%%%%%%%%%%%%%%%%%%%%%%%%%%%%%%%%%%%%
% confluence on normalizing terms
%%%%%%%%%%%%%%%%%%%%%%%%%%%%%%%%%%%%%%%%%%%%%%%%%%%%%%%%%%%%%%%%%%%%%%%%%%%

\subsection{Confluence on weakly $\b$-normalizing terms}
\label{par-Aweak}

We now turn to the problem of dropping the left-linearity and
semi-closure conditions.

%%%%%%%%%%%%%%%%%%%%%%%%%%%%%%%%%%%%%%%%%%%%%%%%%%%%%%%%%%%%%%%%%%%%%%%%%%%
% Preliminaries
%%%%%%%%%%%%%%%%%%%%%%%%%%%%%%%%%%%%%%%%%%%%%%%%%%%%%%%%%%%%%%%%%%%%%%%%%%%

As seen above, fixpoint combinators make the commutation of $\a^*_\b$
and $\a^*_{\cA}$ fail when rewriting involves equality tests
between open terms. When using weakly $\b$-normalizing terms, we can
project rewriting on $\b$-normal forms ($\bnf$), thus eliminating
fixpoints as soon as they are not significant for the reduction.

Hence, we seek to obtain $\bnf(s) \a^*_\cA \bnf(t)$ whenever
$s\a^*_{\b\cup\cA} t$. This requires three important
properties.

First, $\b$-normal forms should be stable by
rewriting. Hence, we assume that right-hand sides are
algebraic.  Moreover, we re-introduce some information from the
algebraic framework, giving maximal arities to function symbols in
$\cF$.

Second, we need normalizing $\b$-derivations to commute with
rewriting. This follows from using the leftmost-outermost strategy of
$\la$-calculus \cite{barendregt84book}.

Finally, we need rule
conditions to be algebraic. 
Indeed, consider the rule
$x\, {\sf b} = y \sgt {\sf f}\,x\,y \a {\sf a}$
that contains an non-algebraic condition.
The relation $\a_\cA$ is confluent but
${\sf a} \al^*_{\b \cup \cA} 
{\sf f}\, (\la x.x)\, ((\la z.z)(\la x.x)\,{\sf b}) 
\a^*_\b {\sf f}\, (\la x.x)\, {\sf b}$
is an unjoinable critical peak.

%%%%%%%%%%%%%%%%%%%%%%%%%%%%%%%%%%%%%%%%%%%%%%%%%%%%%%%%%%%%%%%%%%%%%%%%%%%
% Arity compliance
%%%%%%%%%%%%%%%%%%%%%%%%%%%%%%%%%%%%%%%%%%%%%%%%%%%%%%%%%%%%%%%%%%%%%%%%%%%

\begin{definition}[Arity-compliance]
We assume that every symbol $f \in \cF$ is equipped with an
arity $\alpha_f \ge 0$. A term is {\em arity-compliant} if it
contains no subterm of the form $f\vt$ with $f \in\cF$ and
$|\vt|>\alpha_f$. A rule $\vd=\vc\sgt l\a r$ is {\em almost
arity-compliant} if $l$ and $r$ are arity-compliant and $l$ is of the
form $f \vl$ with $|\vl|=\alpha_f$. A rule is {\em
arity-compliant} if, furthermore, $\vd$ and $\vc$ are arity-compliant.
Let $\cU$ be the set of terms having an arity-compliant $\b$-normal form.
\end{definition}

Remark that a higher-order rule (with active variables and
abstractions) can be arity-compliant.

\comment{
The notion of arity-compliance can be explained as follows:
it prevents a collapsing rule from creating 
a $\b$-redex.
}

Arity-compliance is useful because it prevents collapsing rules
from creating $\b$-redexes. 
For example, the rule
${\sf id}\; x \a x$ forces the arity of {\sf id} to be $1$.  Hence the
term ${\sf id}\, (\la x.x) \, y$ is not arity-compliant.  Moreover it
is a $\b$-normal form that $\a_\cA$-reduces to the $\b$-redex $(\la
x.x)y$. It is then easy to build an arity-uncompliant term that
makes the preservation of confluence to fail. Let $Y=\omega_{\sf s}
\omega_{\sf s}$ with $\omega_{\sf s} = \la x.{\sf s} \, x \, x$. The
term $ - \, ({\sf id}\, \omega_{\sf s} \omega_{\sf s}) \, ({\sf id}\,
\omega_{\sf s} \omega_{\sf s})$ is an arity-uncompliant $\b$-normal
form. Reducing the {\sf id}'s leads to $- \, Y \, Y$ which is the head
of an unjoinable critical peak.

%%%%%%%%%%%%%%%%%%%%%%%%%%%%%%%%%%%%%%%%%%%%%%%%%%%%%%%%%%%%%%%%%%%%%%%%%%%
% weak beta-normalization
%%%%%%%%%%%%%%%%%%%%%%%%%%%%%%%%%%%%%%%%%%%%%%%%%%%%%%%%%%%%%%%%%%%%%%%%%%%

However, we do not assume that every term at hand is arity-compliant.
Indeed, a term that has an arity-compliant $\b$-normal form does not need
to be arity-compliant itself. More precisely, for a weakly
$\b$-normalizing term, the leftmost-outermost strategy (for $\a_\b$)
never evaluates subterms that are not $\b$-normalizing and it follows
that such subterms may be arity-uncompliant without disturbing the
projection on $\b$-normal forms.

The point is the well-foundedness of the leftmost-outermost strategy
for $\a_\b$ on weakly $\b$-normalizing terms
\cite{barendregt84book}. This strategy can be described by means of
{\em head} $\b$-reductions, that are easily shown to commute with
(parallel) conditional rewriting. 
Any $\la$-term can be written $\la \vx.v\, a_0\, a_1 \dots a_n$ where either
$v \in \cX \cup \cF$ (a) or $v$ is a $\la$-abstraction (b).
We denote by $\a_h$ the head $\b$-step 
$\la \vx.(\la y.b) a_0 \va \a_h \la \vx.b\{y\to a_0\} \va$. Let
$s\succ t$ iff either $s$ is of the form (b) and $s\a_h t$, or $s$ is of the form
(a) with $n \geq 1$ and $t = a_i$ for some $i \geq 0$.
In the latter case, the free variables of $t$ can be bound in $s$.
Hence, $t$ can be a subterm of a term $\alpha$-equivalent to $s$ ;
for instance $\la x.{\sf f}x \succ y$ for all $y \in \cX$.

\begin{lemma}\label{lem-succ}
Let $\cW\cN$ be the set of weakly $\b$-normalizing terms ;
(i) if $s \in \cW\cN$ and $s \succ t$ then $t \in \cW\cN$,
(ii) $\succ$ is well-founded on $\cW\cN$.
\end{lemma}

%%%%%%%%%%%%%%%%%%%%%%%%%%%%%%%%%%%%%%%%%%%%%%%%%%%%%%%%%%%%%%%%%%%%%%%%%%%
% Parallelization of conditional rewriting
%%%%%%%%%%%%%%%%%%%%%%%%%%%%%%%%%%%%%%%%%%%%%%%%%%%%%%%%%%%%%%%%%%%%%%%%%%%

It follows that we can reason by well-founded induction on $\succ$. 
For all $i \geq 0$, we
use a nested parallelization of $\a_{\cA_i}$.
It corresponds to the one used in \cite{OR94}, that can be seen as a generalization
of Tait and Martin-L{\"o}f parallel relation.
As for $\tgt_\b$ and $\a_{\b}$, in the orthogonal case,
a complete development of $\a_{\cA_i}$ can be simulated
by {\em one step} $\tgt_{\cA_i}$-reduction.
This relation is also
an adaptation to conditional rewriting of the
parallelization used in \cite{dougherty92ic}.

\begin{definition}[Conditional nested parallel relations]
\label{def-walkA}
For all $i \geq 0$, let $\tgt_{\cA_i}$ be the smallest parallel
rewrite relation closed by:
\[(rule) \quad
\dfrac{\vd = \vc \sgt l \a r \in \cR \quad l\s \a_{\cA_i} r\s \quad \s \tgt_{\cA_i} \t}
	{ l\s \tgt_{\cA_i} r\t }\]
\end{definition}

Recall that $l\s \a_{\cA_i} r\s$ is ensured by $\vd\s \ad_{\cA_{i-1}} \vc\s$. 
These relations enjoy some nice
properties: (1) ${\a_{\cA_i}}\sle{\tgt_{\cA_i}}\sle{\a_{\cA_i}^*}$,
(2) ${s\tgt_{\cA_i} t}$ $\A$ ${u\{x \to s\}}\tgt_{\cA_i}{u\{x\to
t\}}$ and (3) ${[s\tgt_{\cA_i} t ~\&~ u\tgt_{\cA_i} v]}~\A~ {u\{x\to
s\}\tgt_{\cA_i} v\{x\to t\}}$. The last one implies commutation of
$\tgt_{\cA_i}$ and $\a_h$. Commutation of rewriting with head
$\b$-reduction has already been coined in \cite{BFG97:jfp}. We now turn to the main lemma.

\begin{lemma}
\label{lem-projA-bnf}
Let $\cR$ be an arity-compliant algebraic system. If $s\in\cU$ and
$s\a_{\b\cup\cA}^* t$, then $t\in\cU$ and $\bnf(s)\a_{\cA}^*
\bnf(t)$.
\end{lemma}

\begin{prf}
We show by induction on $i$ the property for $\a^*_{\b \cup \cA_i}$.
We denote by (I) the corresponding induction hypothesis.
The case $i=0$ is trivial. Assume that $i > 0$.
An induction on the number of $\a_{\b\cup\cA_i }$-steps leads us to prove that $\bnf(s)
\tgt_{\cA_i} \bnf(t)$ whenever $s \tgt_{\cA_i} t$ and $s$ has an
arity-compliant $\b$-normal form. We reason by induction on
$\succ$.

First (1), assume that $s$ is of the form (a).
If no rule is reduced at its head, the
result follows from induction hypothesis on $\succ$.
Otherwise, there is a rule $\vd = \vc \sgt l \a r$ such that
$s = \la \vx.l\s\va$ and $t = \la \vx.r\t\vb$
with
$l\s \tgt_{\cA_i} r\t$ and $\vd\s \ad_{\cA_{i-1}} \vc\s$.
Since $l$ is algebraic, $\bnf(s)$ is of the form $\la \vx. l\s' \va'$
where $\s' = \bnf(\s)$ and $\va' = \bnf(\va)$. Since $\bnf(s)$ is arity-compliant,
$\va' = \emptyset$, hence $\va = \emptyset$ and $s = \la \vx . l\s$.
Therefore, because $l\s \tgt_{\cA_i} r\t$, we have $\vb = \emptyset$ and 
$t = \la \vx.r\t$.
It remains to show that $t$ has an arity-compliant normal form
and that $\bnf(s) = \la \vx.l\s' \tgt_{\cA_i} \bnf(t)$.
Because $l$ is algebraic, its variables are $\prec^+ l$.
We can then apply induction hypothesis on $\s \tgt_{\cA_i} \t$.
It follows that $\t$ has an arity-compliant normal form $\t'$
with $\s' \tgt_{\cA_i} \t'$.
Since $r$ is algebraic, $\la \vx.r\t'$ is the (arity-compliant)
$\b$-normal form of $t$. 
Hence it remains to show that
$l\s' \tgt_{\cA_i} r\t'$.
Because $\s' \tgt_{\cA_i} \t'$, it suffices to prove that
$l\s' \a_{\cA_i} r\s'$.
Thus, we are done if we show that $\vd\s' \ad_{\cA_{i-1}} \vc\s'$.
Since $\vd$ and $\vc$ are algebraic, $\bnf(\vd\s) = \vd\s'$ and
$\bnf(\vc\s) = \vc\s'$.
Now, since $\vd$ is algebraic and arity-compliant and $\s'$ is arity compliant,
$\vd\s'$ is arity-compliant. The same holds for $\vc\s'$.
Hence we conclude by applying induction hypothesis (I) on
$\vd\s \ad_{\cA_{i-1}} \vc\s$.

Second (2), when $s$ is of the form (b) we head $\b$-normalize it and
obtain a term $s'$ of the form (a) having an arity-compliant $\b$-normal form.
Using commutation of $\tgt_{\cA_i}$ and $\a_h$, 
we obtain a term $t'$ such that $s' \tgt_{\cA_i} t'$.
Since $s \succ^+ s'$, we can reason as in case (1).
\end{prf}

%%%%%%%%%%%%%%%%%%%%%%%%%%%%%%%%%%%%%%%%%%%%%%%%%%%%%%%%%%%%%%%%%%%%%%%%%%%
% confluence out of terms having an AC \bNF
%%%%%%%%%%%%%%%%%%%%%%%%%%%%%%%%%%%%%%%%%%%%%%%%%%%%%%%%%%%%%%%%%%%%%%%%%%%

The preservation of confluence is a direct consequence of the
projection on $\b$-normal forms.

\begin{theorem}
\label{thm-conflA-bnf}
Let $\cR$ be an arity-compliant algebraic system such that $\a_\cA$ is
confluent. Then, $\a_{\b\cup\cA}$ is confluent on $\cU$.
\end{theorem}

%%%%%%%%%%%%%%%%%%%%%%%%%%%%%%%%%%%%%%%%%%%%%%%%%%%%%%%%%%%%%%%%%%%%%%%%%%%
% related work
%%%%%%%%%%%%%%%%%%%%%%%%%%%%%%%%%%%%%%%%%%%%%%%%%%%%%%%%%%%%%%%%%%%%%%%%%%%
\paragraph{Comparison with Dougherty's work.}
This section is an extension of \cite{dougherty92ic}. We give a
further exploration of the idea that preservation of confluence, when
using hypothesis on $\a_\b$, should be independent from any typing
discipline for the $\la$-calculus.

Moreover, we extend its result in three ways.
First, we adapt it to conditional rewriting.
Second, we allow nested symbols in lhs to be applied to less arguments than their arity.
And third, we
use weakly $\b$-normalizing terms whose normal forms are arity-compliant ;
whereas Dougherty uses the set of strongly normalizing
arity-compliant terms which is closed by reduction.

\comment{
In that paper, Dougherty proves, in the non-conditional case, that
left-linearity is not needed when considering so called $\cR$-stable
sets of terms. A set of terms is $\cR$-stable if it is made of
$\b$-strongly normalizing arity-compliant terms and is closed under
$\a_\b$, $\a_\cA$ and subterm. Then, $\a_{\b\cup\cA}$ is confluent on
any $\cR$-stable set of terms whenever $\a_\cA$ is confluent and $\cR$
is first-order.

We extend this result in three ways. First, we adapt it to conditional
rewriting. The main difference is that we need a more subtle induction
relation. Second, we do not limit ourselves to curried versions of
first-order rewrite rules. We allow nested symbols in rules to be
applied to less arguments that their arity. Third, and that is the
main point, we show that confluence is preserved out of terms having
an arity-compliant $\b$-normal form. The amelioration is
two-fold. First we do not need to assume that {\em every}
$\b$-reduction starting from {\em any} $\a_{\b\cup\cA}$-reduct of a
term terminates. We simply need that {\em there exists} a terminating
$\b$-reduction starting from the term. Second, we do not need to
assume that {\em every} $\a_{\b\cup\cA}$-reduct of a term is
arity-compliant but only that its $\b$-normal form is.
}

%%%%%%%%%%%%%%%%%%%%%%%%%%%%%%%%%%%%%%%%%%%%%%%%%%%%%%%%%%%%%%%%%%%%%%%%%%%
% confluence of beta union B
%%%%%%%%%%%%%%%%%%%%%%%%%%%%%%%%%%%%%%%%%%%%%%%%%%%%%%%%%%%%%%%%%%%%%%%%%%%

\section{Using $\a_\b$ in the evaluation of conditions}\label{sec-Bconfl}

The goal of this section is to give conditions on $\cR$ to deduce
confluence of $\a_{\b\cup\cB}$ from confluence of $\a_{\cA}$.
We achieve this by exhibiting two different criteria ensuring that
\begin{equation}
{\a_{\b\cup\cB}^*} \sle {\a_\b^*\a_\cA^*\al_\b^*} ~.
\tag{$\st$}
\end{equation}

The first
case concerns left-linear and semi-closed systems.
This holds only on some sets of terms that, after Dougherty
\cite{dougherty92ic}, we call {\em $\cR$-stable}, although our
definition of stability does not require strong $\b$-normalization
(see Sect.~\ref{par-Aweak} and Def.~\ref{def-stable}). This is an
extra hypothesis compared to the result of Sect.~\ref{par-Alin}. The
second case is a direct extension of Lemma \ref{lem-projA-bnf} to
$\a_{\b\cup\cB}$. In both cases, we assume the rules to be algebraic
and arity-compliant. We are then able to obtain confluence of
$\a_{\b\cup\cB}$ since, in each case, our assumptions ensure that the
results of Sect.~\ref{sec-Aconfl} applies, hence that $\a_{\b\cup\cA}$
is confluent whenever $\a_\cA$ is.

It is important to underline the meaning of $(\st)$. Given an
arity-compliant {\em algebraic} rule $\vd=\vc\sgt l\a r$, every $\b$-redex
occurring in $\vd\s$ or $\vc\s$ also occurs in $l\s$. Then, $(\st)$
means that there is a $\b$-reduction starting from $l\s$ that reduces
these redexes and produce a substitution $\s'$ such that $l\s
\a^*_\b l\s' \a_{\cA} r\s' \al^*_\b r\s$. In other words, if the
conditions are satisfied with $\s$ and $\a_{\b\cup\cB}$ (\ie
$\vd\s\ad_{\b\cup\cB} \vc\s$), then they are satisfied with $\s'$ and
$\a_\cA$ (\ie $\vd\s'\ad_\cA \vc\s'$).
\comment{Recall that non left-linear (or
non semi-closed) rules are dealt with by using the existence and
unicity of a $\b$-normal form for weakly $\b$-normalizing terms.}

We now give some examples of non arity-compliant or non algebraic
rules in which, at the same time, $(\st)$ fails and $\a_{\b\cup\cB}$
is not confluent whereas $\a_{\b\cup\cA}$ for (1), (3), (4) 
and at least $\a_{\cA}$ for (2) is.
\[
\begin{array}{c !{\qquad} l !{\qquad} l !{\qquad} r !{\qquad} l}

(1) & 		 & {\sf g} x \a x {\sf c}           
		 & {\sf g} x  = {\sf d} \sgt {\sf f} x \a {\sf a} 
		 & {\sf f} x \a {\sf b}					\\

(2) & 		 & 	   		
		 & x {\sf c}  = {\sf d} \sgt {\sf f} x \a {\sf a} 
		 & {\sf f} x \a {\sf b}					\\

(3) & 		 & {\sf h} x \a x 	
		 & {\sf h} x {\sf c} = {\sf d} \sgt {\sf f} x \a {\sf a}    
		 & {\sf f} x \a {\sf b} 				\\

(4) & {\sf h} x y \a {\sf g} x y 
		 & {\sf g} x \a x		
		 & {\sf h} x {\sf c} = {\sf d} \sgt {\sf f} x \a {\sf a} 
		 & {\sf f} x \a {\sf b}					\\

\end{array}
\]
The first and second examples respectively contain a rule with a non
algebraic right-hand side and a rule with a non algebraic condition.
Examples $(3)$ and $(4)$ use non arity-compliant terms, in the
conditional part and in the right-hand side of a rule respectively.
For these four examples, the step ${\sf f} (\la x.{\sf d}) \a_{\cB} {\sf a}$
is not in $\a^*_\b \a^*_{\cA} \al^*_\b$ and 
${\sf a} \al_\cB {\sf f} (\la x.{\sf d}) \a_\cB {\sf b}$ is
an unjoinable peak.

However, $(\st)$ is by no means a necessary condition
ensuring that $\a_{\b\cup\cB}$ is confluent when $\a_{\b\cup\cA}$ so
is. In the above examples, confluence of $\a_{\b\cup\cB}$ can be
recovered when adding appropriate rules, yet not restoring $(\st)$.

%% Remarque w.r.t van Oostrom
\comment{
Note that $(\st)$ resembles to a property required on substitution calculus
used in \cite{OR94}.
This would require to see $\a_\b$ as the substitution calculus.
But this does not fit in our framework, in particular because we consider 
$\a_\b$ and rewriting at the same level.
Moreover, the substitution calculus used in \cite{OR94} are required
to be complete (\ie strongly normalizing and confluent),
which may not be the case here for $\a_\b$.}
%% Remarque w.r.t van Oostrom

As we are interested in deducing the confluence of $\a_{\b\cup\cB}$
from the confluence of $\a_\cA$, it is more convenient to take in
Def. \ref{def-cond-rew} $\a_\cB=\bigcup_{i\geq 0}\a_{\cB_i}$ with
$\a_{\cB_0}=\a_\cA$ instead of $\a_{\cB_0}=\emptyset$ (this does not
change $\a_\cB$ since $\a_\cA\sle\a_\cB$).

%%%%%%%%%%%%%%%%%%%%%%%%%%%%%%%%%%%%%%%%%%%%%%%%%%%%%%%%%%%%%%%%%%%%%%%%%%%
% left linear systems
%%%%%%%%%%%%%%%%%%%%%%%%%%%%%%%%%%%%%%%%%%%%%%%%%%%%%%%%%%%%%%%%%%%%%%%%%%%

\subsection{Confluence of left-linear systems}\label{par-Blin}

%%%%%%%%%%%%%%%%%%%%%%%%%%%%%%%%%%%%%%%%%%%%%%%%%%%%%%%%%%%%%%%%%%%%%%%%%%%
% beta-reduction from algebraic cap
%%%%%%%%%%%%%%%%%%%%%%%%%%%%%%%%%%%%%%%%%%%%%%%%%%%%%%%%%%%%%%%%%%%%%%%%%%%

In this paragraph, we prove $(\st)$ provided that rules are
arity-compliant, algebraic, left-linear and semi-closed. This
inclusion is shown on {\em $\cR$-stable} sets of terms.

\begin{definition}[$\cR$-stable sets]
\label{def-stable}
Let $\cR$ be a set of rules. A set $\cS$ is {\em almost
$\cR$-stable} if it contains only arity-compliant terms, is stable by 
subterm and $\b$-reduction, and $C[r\s]\in\cS$ whenever $C[l\s] \in \cS$
and $\vd = \vc \sgt l \a r \in \cR$.
An almost $\cR$-stable set $\cS$ is $\cR$-stable if 
$\vd\s,\vc\s \in \cS$ whenever $C[l\s] \in \cS$ and
$\vd = \vc \sgt l \a r \in \cR$.
\end{definition}

This includes the set of strongly $\a_{\b\cup\cA}$-normalizable
arity-compliant terms and any of its subset closed by subterm and
reduction, by using a simple type discipline for instance.

The inclusion $(\st)$ is proved by induction on the stratification of
$\a_{\cB}$ with $\a_{\cB_0}=\a_\cA$. The base case corresponds to
${\a_{\b\cup\cA}^*}\sle {\a_\b^*\a_\cA^*\al_\b^*}$, which does not require
rule conditions to be algebraic nor arity-compliant.

The previous examples show however that this may fail in presence of
arity-uncompliant or non-algebraic right-hand sides. Note that the
result is proved only on almost $\cR$-stable sets of terms. Note also that a
set containing a term reducible by the first rule of example $(4)$
above is obviously not stable. Finally, note that the $\b$-expansion
steps are needed because rules can be duplicating.

%%%%%%%%%%%%%%%%%%%%%%%%%%%%%%%%%%%%%%%%%%%%%%%%%%%%%%%%%%%%%%%%%%%%%%%%%%%
% postponement with several steps
%%%%%%%%%%%%%%%%%%%%%%%%%%%%%%%%%%%%%%%%%%%%%%%%%%%%%%%%%%%%%%%%%%%%%%%%%%%

\begin{lemma}
\label{lem-post-star}
Let $\cR$ be a semi-closed left-linear right-algebraic system.
On any almost $\cR$-stable set of terms,
${\a_{\b\cup\cA}^*}\sle {\a_\b^*\a_\cA^*\al_\b^*}$.
\end{lemma}

\begin{prf}
The proof is in four steps.  We begin (1) to show that $\a_\cA \tgt_\b
\sle \tgt_\b \a^*_{\cA} \tlt_\b$, reasoning by cases on the step
$\tgt_\b$. This inclusion relies on an important fact of algebraic
terms: if $s$ is algebraic and $s\s\tgt_\b v$ then $v \tgt_\b s\s'$
with $\s \tgt^*_\b \s'$.  From (1), it follows that (2)
$\a_\cA^*\tgt_\b\sle \tgt_\b\a_\cA^*\tlt_\b^*$, by induction on the
number of $\a_\cA$-steps. Then (3), we obtain $\a_\cA^*\tgt_\b^*\sle
\tgt_\b^*\a_\cA^*\tlt_\b^*$ using an induction on the number of
$\tgt_\b$-steps and the diamond property of $\tgt_\b$.
Finally (4), we deduce that $(\tgt_\b\cup\a_\cA)^*\sle
\tgt_\b^*\a_\cA^*\tlt_\b^*$ by induction on the length of
$(\tgt_\b\cup\a_\cA)^*$.
\end{prf}

%%%%%%%%%%%%%%%%%%%%%%%%%%%%%%%%%%%%%%%%%%%%%%%%%%%%%%%%%%%%%%%%%%%%%%%%%%%
% from B to A
%%%%%%%%%%%%%%%%%%%%%%%%%%%%%%%%%%%%%%%%%%%%%%%%%%%%%%%%%%%%%%%%%%%%%%%%%%%

We now turn to the main result of this subsection. As seen in the
previous examples, rules have to be algebraic and arity-compliant.

\begin{lemma}
\label{lem-BtoA}
Let $\cR$ be a semi-closed left-linear algebraic
system. On any $\cR$-stable set of terms, ${\a_{\b\cup\cB}^*}\sle
{\a_\b^*\a_\cA^*\al_\b^*}$.
\end{lemma}

\begin{prf}
The first point is to see that (1) $\a_{\cB_1}^*\sle
\a_\b^*\a_\cA^*\al_\b^*$. This is done by induction on the number of
$\cB_1$-steps, using Lemma \ref{lem-post-star}. We then deduce (2)
$\a_{\b\cup\cB_1}^*\sle \a_\b^*\a_\cA^*\al_\b^*$, by induction on the
number of $\a_{\b\cup\cB_1}$-steps.
The result follows from an induction on $i$ showing that
$\a_{\cB_i}\sle \a_{\cB_1}$.
\end{prf}

%%%%%%%%%%%%%%%%%%%%%%%%%%%%%%%%%%%%%%%%%%%%%%%%%%%%%%%%%%%%%%%%%%%%%%%%%%%
% confluence of beta union \cB
%%%%%%%%%%%%%%%%%%%%%%%%%%%%%%%%%%%%%%%%%%%%%%%%%%%%%%%%%%%%%%%%%%%%%%%%%%%
\begin{theorem}
\label{thm-Bconfl}
Assume that $\cR$ is a semi-closed left-linear
algebraic system. If $\a_\cA$ is confluent, then $\a_{\b\cup\cB}$ is
confluent on any $\cR$-stable set of terms.
\end{theorem}

Recall that in this case $\a_{\b \cup \cA}$-confluence follows from $\a_{\cA}$-confluence
by Thm. \ref{thm-conflA}.

%%%%%%%%%%%%%%%%%%%%%%%%%%%%%%%%%%%%%%%%%%%%%%%%%%%%%%%%%%%%%%%%%%%%%%%%%%%
% confluence on stable sets of strongly $\b$-normalizing terms
%%%%%%%%%%%%%%%%%%%%%%%%%%%%%%%%%%%%%%%%%%%%%%%%%%%%%%%%%%%%%%%%%%%%%%%%%%%

\subsection{Confluence on weakly $\b$-normalizing terms}
\label{par-Bweak}

This subsection concerns the straightforward extension to $\a_\cB$ of
the results of Sect.~\ref{par-Aweak}. The definition of $\tgt_{\cB_i}$
follows the same scheme as the one of $\tgt_{\cA_i}$; the only
difference is that $\cB_i$ is used everywhere in place of $\cA_i$.  It
follows that given a rule $\vd=\vc\sgt l\a r$, to have
$l\s\tgt_{\cB_i} r\t$, we must have $\s\tgt_{\cB_i} \t$ and
$\vd\s\ad_{\b\cup\cB_{i-1}} \vc\s$. The relations $\tgt_{\cB_i}$ enjoy
the same nice properties as the $\tgt_{\cA_i}$'s.

\begin{lemma}
\label{lem-projB-bnf}
Let $\cR$ be an arity-compliant algebraic system. If $s\in\cU$ and
$s\a_{\b\cup\cB}^* t$, then $t\in\cU$ and $\bnf(s)\a_{\cA}^*
\bnf(t)$.
\end{lemma}

The only difference in the proof is that the case $i = 0$ is now
ensured by Lemma \ref{lem-projA-bnf} (since $\a_{\cB_0}=\a_{\cA}$).
The theorem follows easily:

\begin{theorem}
\label{thm-conflB-bnf}
Let $\cR$ be an arity-compliant algebraic system such that $\a_\cA$ is
confluent. Then, $\a_{\b\cup\cB}$ is confluent on $\cU$.
\end{theorem}

%%%%%%%%%%%%%%%%%%%%%%%%%%%%%%%%%%%%%%%%%%%%%%%%%%%%%%%%%%%%%%%%%%%%%%%%%%%
% orthonormal systems
%%%%%%%%%%%%%%%%%%%%%%%%%%%%%%%%%%%%%%%%%%%%%%%%%%%%%%%%%%%%%%%%%%%%%%%%%%%

\section{Orthonormal systems}\label{sec-exclus}

In this section, we give a criterion ensuring confluence of
$\a_{\b\cup\cB}$ when conditions and right-hand sides possibly contain
abstractions and active variables.

This criterion comes from peculiarities of orthogonality with
conditional rewriting. In non-conditional rewriting, a system is
orthogonal when it is left-linear and has no critical pair. A critical
pair comes from the superposition of two different rule left-hand
sides at non-variable positions. The general definition of orthogonal
conditional systems is the same.  But, in conditional rewriting, there
can be superpositions of two different rules left-hand sides whose
conditions cannot be satisfied with the same substitution.  Such
critical pairs are said {\em infeasible} and it could be profitable to
consider systems whose critical pairs are all infeasible.
 
In \cite{ohlebusch02book}, it is remarked that results on the
confluence of natural and normal orthogonal conditional systems should
be extended to systems that have no feasible critical pair. But the
results obtained this way are not directly applicable since proving
unfeasibility of critical pairs may require confluence.
In Takahashi's work \cite{taka93tlca}, conditions can be any predicate $P$ on terms.
Confluence is proved with the assumption that they are stable by reduction:
if $P\s$ holds and $\s \a \t$, then $P\t$ holds.
For the systems studied in this section, stability of conditions by
reduction precisely follows from confluence. Hence the results of~\cite{taka93tlca}
do not directly apply.

The purpose of this section is to give a {\em syntactic} condition on rules that imply
unfeasibility of critical pairs, hence confluence.

%%%%%%%%%%%%%%%%%%%%%%%%%%%%%%%%%%%%%%%%%%%%%%%%%%%%%%%%%%%%%%%%%%%%%%%%%%%
% critical pairs
%%%%%%%%%%%%%%%%%%%%%%%%%%%%%%%%%%%%%%%%%%%%%%%%%%%%%%%%%%%%%%%%%%%%%%%%%%%

\begin{definition}[Conditional critical pairs]
Given two rules $\vd=\vc\sgt l\a r$ and $\vd'=\vc'\sgt l'\a r'$,
if $p$ is a non-variable position in $l$ and $\s$ is a most general
unifier of $l|_p$ and $l'$, then
\[
\vd\s=\vc\s \land \vd'\s=\vc'\s \sgt (l[r']_p\s, r\s)
\]
is a {\em conditional critical pair}. A critical pair $\vd=\vc\sgt
(s,t)$ is {\em feasible} for $\a_\cA$ (resp. $\a_{\cB}$) if there is a
substitution $\s$ such that $\vd\s\ad_{\cA} \vc\s$
(resp. $\vd\s\ad_{\b\cup\cB} \vc\s$).
\end{definition}

As an example, consider the rules used to define {\sf occ} in
Sect.~\ref{sec-ex}. There is a superposition between the left-hand
sides of the last two rules giving the critical peak ${\sf ff}
\al {\sf occ}\, (x :: o)\, ({\sf nd}\, y\, l) \a {\sf occ}\, o\,
({\sf get}\, l \, x)$. But a peak of this form can occur only if there
are two terms $s,t$ such that ${\sf tt} \:\al^*\: \geq\, ({\sf len}\,
s)\, t \:\a^*\: {\sf ff}$.  Using the stratification of $\a_\cA$, the
confluence of $\a_{\cA_i}$ implies that this pair is not
feasible. Hence the above peak cannot occur with $\a_{\cA_{i+1}}$ and
this relation is confluent.

This method can be used on systems with higher-order terms in
right-hand sides and conditions, as for example the rules defining
{\sf app} and {\sf filter}. Hence, it is useful for proving the
confluence of $\a_{\b\cup\cB}$ for systems where this relation does not need to be
included in $\alr^*_{\b\cup\cA}$. In this section, we
generalize the method and apply it on a class of systems called {\em
orthonormal}. As in the previous section, we use stratification of
$\a_\cB$, but now with $\a_{\cB_0}=\emptyset$. A symbol $f \in\cF$ 
is {\em defined} if it is the head of a rule left-hand side.

%%%%%%%%%%%%%%%%%%%%%%%%%%%%%%%%%%%%%%%%%%%%%%%%%%%%%%%%%%%%%%%%%%%%%%%%%%%
% orthonormal systems
%%%%%%%%%%%%%%%%%%%%%%%%%%%%%%%%%%%%%%%%%%%%%%%%%%%%%%%%%%%%%%%%%%%%%%%%%%%

\begin{definition}[Orthonormal systems]
A system is {\em orthonormal} if (1) it is left-linear; (2) in every
rule $\vd=\vc\sgt l\a r$, the terms in $\vc$ are closed $\b$-normal
forms not containing defined symbols; and (3) for every critical pair
$\vd=\vc\sgt (s,t)$, there exists $i\neq j$ such that
$d_i=d_j$ and $c_i\neq c_j$.
\end{definition}

Note that an orthonormal system is left-linear and semi-closed, but
does not need to be arity-compliant or algebraic. Note also that the
form of the conditions leads to a {\em normal} conditional rewrite
relation. The reader can check that the whole system given in
Sect.~\ref{sec-ex} is orthonormal.

%%%%%%%%%%%%%%%%%%%%%%%%%%%%%%%%%%%%%%%%%%%%%%%%%%%%%%%%%%%%%%%%%%%%%%%%%%%
% Shallow confluence
%%%%%%%%%%%%%%%%%%%%%%%%%%%%%%%%%%%%%%%%%%%%%%%%%%%%%%%%%%%%%%%%%%%%%%%%%%%

We now prove that $\a_{\b\cup\cB}$ is shallow confluent 
(\ie $\a^*_{\b \cup \cB_i}$ and $\a^*_{\b \cup \cB_j}$ commute for all $i,j\ge 0$)
when $\cR$ is
orthonormal. The first point is that confluence of $\a_{\b \cup \cB_i}$ implies
commutation of $\a^*_\b$ and $\a^*_{\cB_{i+1}}$.
The proof is as in Sect.~\ref{par-Alin}, except that in a rule $\vd = \vc \sgt l \a r$,
$\vc$ are closed $\a_{\b \cup \cB}$-normal forms.
The main Lemma concerns commutation of parallel relations
of $\tgt_{\cB_i}$ and $\tgt_{\cB_j}$ for all $i,j \geq 0$.
But here, we use a weak form of parallelization: $\tgt_{\cB_i}$ is simply
the parallel closure of $\a_{\cB_i}$.
The name of the Lemma is usual for this kind of result with rewriting 
(see \cite{ohlebusch02book}).
Write $<_{mul}$ for the multiset extension of the usual ordering on naturals numbers.

\begin{lemma}[Parallel Moves]\label{lem-par-moves}
Let $\cR$ be an orthonormal system.
If $\{n,m\} <_{mul} \{i,j\}$ implies commutation of $\a^*_{\b \cup \cB_n}$ and
$\a^*_{\b \cup \cB_m}$, then $\tgt_{\cB_i}$ and $\tgt_{\cB_j}$ commute.
\end{lemma}

\begin{prf}
The key point is the commutation of $\a^*_{\b \cup \cB_n}$ and
$\a^*_{\b \cup \cB_m}$ for $\{n,m\} <_{mul} \{i,j\}$.  It implies that
two rules whose respectives conditions are satisfied with $\a^*_{\b
\cup \cB_i}$ and $\a^*_{\b \cup \cB_j}$ are not superposable at
non-variable positions.  The rest of the proof follows usual schemes
(see Sect. 7.4 in \cite{ohlebusch02book}).
\end{prf}

Now, an induction on $<_{mul}$ provides the commutation of
$\a_{\b\cup\cB_i}$ and $\a_{\b\cup\cB_j}$ for all $i,j\geq 0$. Shallow
confluence immediately follows.

\begin{theorem}
\label{thm:lcr:beta:cond}
If $\cR$ is an orthonormal system, then $\a_{\b\cup\cB}$ is shallow
confluent.
\end{theorem}

Hence, the relation $\a_{\b\cup\cB}$ induced by the system of Sect.~\ref{sec-ex}
is confluent.

%%%%%%%%%%%%%%%%%%%%%%%%%%%%%%%%%%%%%%%%%%%%%%%%%%%%%%%%%%%%%%%%%%%%%%%%%%%
% conclusion
%%%%%%%%%%%%%%%%%%%%%%%%%%%%%%%%%%%%%%%%%%%%%%%%%%%%%%%%%%%%%%%%%%%%%%%%%%%

\section{Conclusion}
%\comment{
Our results are summarized in the following table.

\newcommand{\acalg}{\begin{tabular}{c} Arity-Compliant \\ \& Algebraic \end{tabular}}
\newcommand{\scalg}{\begin{tabular}{c} Semi-Closed \\ \& Algebraic \end{tabular}}

\begin{center}
\begin{tabular}
{| c     	| c 		| c 		| c 		| c 		| c |}
\hline
\S	& Terms 	& Lhs 		& Rhs 		& Conditions 	& Result	\\
\hline
\hline

\ref{par-Alin}  	& $\cT$		& Linear 	& Applicative 	& Semi-Closed 	& 
		\begin{tabular}{c}
		$\a_\cA$ Confluent $\A$ \\ $\a_{\cA \cup \b}$ Confluent
		\end{tabular} \\
		% $\a_\cA$ CR $\A$ $\a_{\cA \cup \b}$ CR \\
\hline		
\ref{par-Aweak}	& $\cU$ 	& 		& \acalg 	&  \acalg & idem \\
\hline
\hline
\ref{par-Blin}	& $\cR$-stable  & Linear 	& Algebraic 	& \scalg 	& 
		\begin{tabular}{c}		
		$\a_\cA$ Confluent $\A$ \\ $\a_{\cB \cup \b}$ Confluent 
		\end{tabular} \\
\hline
\ref{par-Bweak}	& $\cU$ 	& 		& \acalg & \acalg & idem \\
\hline
\hline
\ref{sec-exclus}& $\cT$		& Linear 	& 		& Orthonormal 	&
		\begin{tabular}{c}
		$\a_{\cB \cup \b}$ \\ Shallow Confluent \\
		\end{tabular}\\
\hline
\end{tabular}
\end{center}
%}

We provide detailed conditions to ensure modularity of confluence when
combining $\b$-reduction and conditional rewriting, either when the
evaluation of conditions uses $\b$-reduction or when it does not.
This has useful applications on the high-level specification side and
for enriching the conversion used in logical frameworks or proof
assistants, while still preserving the confluence property.

These results lead us to the following remarks and further research points. The results
obtained in Sect.~\ref{sec-Aconfl} and~\ref{sec-Bconfl} for the standard conditional
rewrite systems extend to the case of oriented systems (hence to normal systems) and to
the case of level-confluent natural systems.  For natural systems, the proofs follow the
same scheme, provided that level-confluence of $\a_\cA$ is assumed. However, it would be
interesting to know if this restriction can be dropped.

Problems arising from non left-linear rewriting are directly transposed to left-linear
conditional rewriting. The semi-closure condition is sufficient to avoid this, and it
provides the counter part of left-linearity for unconditional rewriting. As a matter of a
fact, it is well known that orthogonal standard conditional rewrite systems are not
confluent, but confluence of orthogonal semi-closed standard systems holds. However, two
remarks have to be made about this restriction. First, it would be interesting to know if
it is a necessary condition and besides, to characterize a class of non semi-closed
systems that can be translated into equivalent semi-closed ones. Second, semi-closed
terminating standard systems behave like normal systems. But normal systems can be easily
translated in equivalent non-conditional systems. Moreover such a translation preserves
good properties such as left-linearity and non-ambiguity. As many of practical uses of
rewriting rely on terminating systems, semi-closed standard systems may be in practice
essentially an intuitive way to design rewrite systems that can be then efficiently
implemented by non-conditional rewriting.

An interesting extension of this work consists in adapting to conditional 
rewriting the axiomatization
and the results of \cite{OR94}. This should leads to a generalization
of the higher-order conditional systems of \cite{avenhaus94HOcondRew}.

{\em Acknowledgments.} We are quite grateful to the anonymous referees for their
constructive and accurate comments and suggestions.

\bibliographystyle{plain}
\bibliography{bibliographie}

%%%%%%%%%%%%%%%%%%%%%%%%%%%%%%%%%%%%%%%%%%%%%%%%%%%%%%%%%%%%%%%%%%%%%%%%%%%
% Proofs of section 3.1
%%%%%%%%%%%%%%%%%%%%%%%%%%%%%%%%%%%%%%%%%%%%%%%%%%%%%%%%%%%%%%%%%%%%%%%%%%%

\section{Proofs of Section~\ref{par-Aweak}}

We begin by proving the well-foundedness of $\succ$.

\begin{lemma}
Let $\cW\cN$ be the set of weakly $\b$-normalizing terms. The set
$\cW\cN$ is stable by $\succ$ and $\succ$ is well-founded on
$\cW\cN$.
\end{lemma}

\begin{proof}
For the first part, let be $s \in \cW\cN$ and $s \succ t$.
If $s$ is of the form (b), the first step of the leftmost-outermost derivation normalizing
$s$ is $t$. Hence $t \in \cW\cN$.
Otherwise, if $t$ has no $\b$-normal form,
then $s$ has no $\b$-normal form.

For the second part, we write $\#(s)$ for the number
of $\a_h$-steps in the leftmost-outermost derivation
starting from $s$ and $|s|$ for the size of $s$.
We show that if $s \succ t$, then $(\#(s),|s|) >_{lex} (\#(t),|t|)$.
If $s$ is of the form (b), by the first point $t \in \cW\cN$.
Since $s \a_h t$, we have $\#(s) > \#(t)$.
Otherwise, the leftmost-outermost strategy starting from $s$
reduces by leftmost-outermost reductions each $a_i$ ($1 \leq i \leq n$).
Hence $\#(s) \ge \#(t)$.
But in this case, $t$ is a proper subterm of $s$, hence $|s| > |t|$.
\end{proof}

Then, we consider the properties (1)-(3) of the walk relations
$\tgt_{\cA_i}$.

\begin{proposition}\label{prop-tgtA}
  For all $i \geq 0$, 
  \begin{enumerate}
  \item $\a_{\cA_i}  ~\subset~  \tgt_{\cA_i} 
    	 ~\subset~ \a_{\cA_i}^*$.
	\label{subset-tgtA}

  \item ${s \tgt_{\cA_i} t} ~\A~ {u\{x \to s\}
    	 \tgt_{\cA_i} u\{x \to t\}}$.
   	\label{tgtA-subst}

  \item ${[s \tgt_{\cA_i} t ~\&~ u \tgt_{\cA_i} v]} ~\A~ {u\{x
  \to s\} \tgt_{\cA_i} v\{x \to t\}}$. 
    \label{tgtA-subst2}
  \end{enumerate}
\end{proposition}

\begin{proof}
The first point is shown by induction on the definition of $\tgt_{\cA_i}$ ; 
the second by induction on $u$.
For the last one, we also use an induction on $\tgt_{\cA_i}$ in $u \tgt_{\cA_i} v$.
If $u$ is $v$, the result is trivial.
If $u \tgt_{\cA_i} v$ was obtained by parallel application or if $u$ is an abstraction,
the result follows from induction hypothesis.
Otherwise, $u \tgt_{\cA_i} v$ is obtained by $(rule)$. That is, there is 
a rule $\vd = \vc \sgt l \a r \in \cR$ such that $u = l\s$, $v = r\t$,
$\s \tgt_{\cA_i} \t$ and $l\s \a_{\cA_i} r\s$.
Since $\a_{\cA_i}$ is a rewrite relation, we have $l\s\{x \to s\} \tgt_{\cA_i} r\s\{x \to s\}$.
By induction hypothesis, we have $\s\{x \to s\} \tgt_{\cA_i} \t\{x \to t\}$.
Therefore $l\s\{x \to s\} \tgt_{\cA_i} r\t\{x \to t\}$.
\end{proof}

We now turn to the commutation of $\tgt_{\cA_i}$ and $\a_h$.
This is a direct consequence of the case (3) of the above Proposition.

\begin{lemma}\label{lem-commut-h}
For all $i \geq 0$, $\tgt_{\cA_i}$ commutes with $\a_h$.
\end{lemma}

\begin{proof}
Assume that 
$s \al_h \la \vx. (\la y.a_0)a_1 \dots a_p \tgt_{\cA_i} t$.
Because rules have non-variable algebraic left hand-sides, 
$t= \la \vx. (\la y.b_0)b_1 \dots b_p$ with
for all $k \in \{0,\dots,p \}$, $a_k
\tgt_{\cA_i} b_k$.
On the other hand, $s = \la \vx. a_0\{y \to a_1\}a_2 \dots a_p$.
It follows from Prop. \ref{prop-tgtA}.\ref{tgtA-subst2} that $a_0\{x \to a_1\}
\tgt_{\cA_i} b_0\{x \to b_1 \}$ (in {\em one} step). 
Hence we have 
$s \tgt_{\cA_i} \la \vx. b_0\{y \to b_1\}b_2 \dots b_p \al_h t$.
\end{proof}

u%%%%%%%%%%%%%%%%%%%%%%%%%%%%%%%%%%%%%%%%%%%%%%%%%%%%%%%%%%%%%%%%%%%%%%%%%%
% Proofs of section 4.1
%%%%%%%%%%%%%%%%%%%%%%%%%%%%%%%%%%%%%%%%%%%%%%%%%%%%%%%%%%%%%%%%%%%%%%%%%%%
\section{Proofs of Section \ref{par-Blin}}

We begin by two technical properties.
The first one is a generalization of the diamond property of
$\tgt_\b$.

\begin{proposition}\label{prop-diamond}
Let be $n \geq 0$ and assume that $s,s_1,\dots, s_n$ are terms such
that for all 
$1 \leq i \leq n$, $s \tgt_\b s_i$.
Then there is a term $s'$ such that for all $1 \leq i \leq n$, $s_i
\tgt_\b s'$.
\end{proposition}

\begin{proof}
The proof is by induction on the structure of $s$.
First, if $s$ is a constant symbol or a variable, then it is a $\b$-normal
from and we are done.
If $s$ is an abstraction $\la x.t$,
then for all $1 \leq i \leq n$, $s_i$ is of the form $\la x.t_i$ and
we conclude by induction hypothesis on $t,t_1,\dots,t_n$.
Now assume that $s$ is an application. There are two cases.
First, $s = t u$ where $t$ is not an abstraction.
Then, for all $1 \leq i \leq n$, $s_i$ is of
the form $t_i u_i$ with $t \tgt_\b t_i$ and $u \tgt_\b u_i$ and we
conclude by induction hypothesis.
Otherwise, $s$ must be of the form $(\la x.t)u$ and for all $1 \leq i
\leq n$, $s_i$ is either of 
the form $(\la x.t_i) u_i$ (1) or of the form $t_i\{x \to u_i \}$ (2).
In both cases we have $t \tgt_\b t_i$ and $u \tgt_\b u_i$. 
By induction hypothesis, there are two terms
$t',u'$ such that for all $1 \leq i \leq n$, $t_i \tgt_\b t'$ and $u_i
\tgt_\b u'$.  
Therefore, in case (1), we have $(\la x.t_i) u_i ~\tgt_\b~ t'\{x \to
u'\}$ and in case (2) $t_i\{x \to u_i \} ~\tgt_\b~ t'\{x \to u'\}$.
Hence, for all $1 \leq i \leq n$, $s_i \tgt_\b t'\{x \to u'\}$.
\end{proof}

For the following Proposition, we define ${\cO}(t,u)$, the 
{\em set of occurrences of $t$ in $u$} as :
${\cO}(t,u) = \vep$ if $t = u$ ; otherwise
${\cO}(t , u_1 u_2) = 1.{\cO}(t,u_1) \cup 2.{\cO}(t,u_2)$,
${\cO}(t , \la x.u) = 1.{\cO}(t,u)$ and ${\cO}(t,x) = {\cO}(t,{\sf f}) = \vide$.

\begin{proposition}
\label{prop-beta-alg}
Let $s$ be an algebraic term.
\begin{enumerate}
\item If $s\s\tgt_\b v$ then there is $\s'$ such that
  $\s\tgt_\b^* \s'$ and $v\tgt_\b s\s'$.
\item If $s\s\tgt_\b^* v$ then there is $\s'$ such that
  $\s\tgt_\b^* \s'$ and $v\tgt_\b^* s\s'$.
\end{enumerate}
\end{proposition}

Note that $s$ does not needs to be linear.

\begin{proof}
  $1.$ Since $s$ is algebraic, every occurrence of $\b$-redex of $s\s$ is of the form
	$p.d$ where $p$ is the occurrence of a variable $x$ in $s$ and 
	$d$ is an occurrence in $\s(x)$.
	Therefore, $v$ is of the form 
	\[
	s[t(x,p)]_{\{ p \tq p \in {\cO}(x,s) ~\&~ x \in FV(s) \}}
	\]
	where, for all $x \in FV(s)$, for all $p \in {\cO}(x,s)$,
	$\s(x) \tgt_\b t(x,p)$.
	By Prop. \ref{prop-diamond}, for all $x \in FV(s)$,
	there exists $t(x)$ such that for all $p \in {\cO}(x,s)$, $t(x,p) \tgt_\b t(x)$.
	Hence 
	\[
	v \tgt_\b s[t(x)]_{\{ p \tq p \in {\cO}(x,s) ~\&~ x \in FV(s) \}} ~.
	\]
	That is, $v \tgt_\b s\s'$ with $\s'(x) = t(x)$.

  $2.$ We reason by induction on the
  number of $\tgt_\b$-steps.
  If $s\s=v$ the result is trivial.
  Otherwise, $s\s \tgt_\b^* v$ is $s\s \tgt_\b v' \tgt_\b^* v$.
  By (1), there is a substitution $\s'$ such that $s\s \tgt_\b v'
  \tgt_\b s\s'$. By strong confluence of $\tgt_\b$, there is a $v''$
  such that $s\s' \tgt_\b^* v'' \tlt_\b^* v$ and the length of $s\s'
  \tgt_\b^* v''$ is no more than the length of $v' \tgt_\b^* v$.
  Hence, we can apply induction hypothesis on $s\s'
  \tgt_\b^* v''$ and thus obtain $\s''$ such that $s\s \tgt_\b^* v
  \tgt_\b^* s\s''$.
\end{proof}

We now turn to the inclusion $\a^*_{\b \cA} \sle \a^*_\b \a^*_\cA \al^*_\b$
on almost $\cR$-stable terms.
We begin by showing that $\a_\cA \tgt_\b \sle \tgt_\b \a^*_\cA \tlt_\b$.

\begin{proposition}
\label{prop-post}
Let $\cR$ be a semi-closed left-linear right-algebraic system.
On any almost $\cR$-stable set of terms,
$\a_\cA \tgt_\b \sle \tgt_\b \a^*_\cA \tlt_\b$.
\end{proposition}

\begin{proof}
Let $R$ be the binary relation be such that, for all $t,u$,
\[
R(t,u) \Alr
\all s~
[~
s \a_\cA t \tgt_\b u ~\A~
\ex s'\,t' 
(s \tgt_\b s' \a^*_\cA t' \tlt_\b t)
~]
\]
We have to show that $R$ is reflexive and compatible with
terms formations rules, parallel application and with the rule $(beta)$.

Reflexivity of $R$ is trivial.
We now prove that $R$ is compatible with term-formation rules,
parallel application and $(beta)$.
\begin{description}
\item[Term-Formation] 
Note that compatibility with parallel application contains
compatibility with application.
Hence compatibility with context is only compatibility with
$\la$-abstraction.

We have to show that if $R(t_1,u_1)$ holds, then $R(\la x.t_1,\la x.u_1)$ holds whenever 
$t_1 \tgt_\b u_1$.
So assume $R(t_1,u_1)$, $t_1 \tgt_\b u_1$ 
and let $s$ be such that $s \a_\cA \la x.t_1 \tgt_\b \la x.u_1$.
Write $t$ for $\la x.t_1$ and $u$ for $\la x.u_1$.
If the contractum of the step $s \a_\cA t$
is in a proper subterm of $t$, we have $s = \la x. s_1$ with 
    $s_1 \a_\cA t_1$ and we conclude by assumption and context compatibility
of $\a_\cA$ and $\tgt_\b$.
Otherwise, there is a rule $\vd = \vc \sgt l \a r$ such that
$s = l\s$ and $t = r\s$.
As $r$ is algebraic, by Prop.~\ref{prop-beta-alg} there is a
  substitution $\s'$ such that $\s \tgt_\b^* \s'$ and $u \tgt_\b r\s'$.
    By linearity of $l$, we have $l\s \tgt_\b l\s'$.
We now show that $l\s' \a_\cA r\s'$. To this end he have to show that
  $\vd\s' \ad_{\cA} \vc$. But $\vd\s \tgt_\b^* \vd\s'$ and by
  Lem.~\ref{lem-commut} there are terms $\vv$ such that $\vd\s'
  \a_{\cA}^* \vv \al_{\b}^* \al_{\cA}^* \vc$.
  Because terms in $\vc$ and right-hand sides of rules
  are build without abstraction symbols, there cannot be any $\b$-step
  starting from an 
  $\a_\cA$-reduct of $\vc$. Hence $\vd\s' \ad_{\cA} \vc$ and we are done.

\item[Parallel application] We have to show that
	\[
	[~ R(t_1,u_1) ~\&~ R(t_2,u_2) ~] ~\A~ R(t_1\,t_2 , u_1\,u_2)
	\]
	whenever $t_1 \tgt_\b u_1$ and $t_2 \tgt_\b u_2$.
	So, assume $R(t_1,u_1)$, $R(t_2,u_2)$, and let
	$s$ be such that $s \a_\cA t_1\,t_2 \tgt_\b u_1\,u_2$
	where $t_i \tgt_\b u_i$.
	Write $t$ for $t_1\, t_2$ and $u$ for $u_1 \, u_2$.
  If the contractum of the step $s \a_\cA t$
is in a proper subterm of $t$ we can conclude by assumption and
context compatibility of $\a_\cA$ and $\tgt_\b$.
Otherwise $s = l\s$ and 
    $t = r\s$ for a rule $\vd = \vc \sgt l \a r$ and we conclude as
  in Case 1.

\item[(beta) rule] 
	We have to show that
	\[
	[~ R(t_1,u_1) ~\&~ R(t_2,u_2) ~] ~\A~ R((\la x.t_1)t_2 , u_1\{x \to u_2\})
	\]
	whenever $t_1 \tgt_\b u_1$ and $t_2 \tgt_\b u_2$.
	So assume $R(t_1,u_1)$, $R(t_2,u_2)$, and let $s$
	be such that $s \a_\cA (\la x.t_1)t_2 \tgt_\b u_1\{x \to u_2\}$
	where $t_i \tgt_\b u_i$.
	Write $t$ for $(\la x.t_1)t_2$ and $u$ for $u_1\{x \to u_2\}$.
	As above, if $s \a_{\cA} t$, is a rooted rewrite
  step, we refer to the Case~1. \\
  Otherwise, as $s$ is arity-compliant, $\la x.t_1$ is not the
  instantiated right hand side of a rule $\vd = \vc \sgt l \a r$.
  Indeed, if it where, we would have $s = f \: \vl \: s_2$ 
  with $l = f \: \vl$. 
  But the term $f \: \vl \: s_2$
  is not arity-compliant, contradicting the hypothesis of stability.
  So we are in cases where $s = (\la x.s_1) t_2$ (resp. $(\la x.t_1)
  s_2$) with $s_1 \a_{\cA} t_1$ (resp. $s_2 \a_{\cA} t_2$). 
  In both cases, we conclude by assumption and context compatibility of
  $\a_\cA$ and $\tgt_\b$.
\end{description}

\end{proof}

\begin{lemma}
\label{lem-post-star-proof}
Let $\cR$ be a semi-closed left-linear right-algebraic system.
On any almost $\cR$-stable set of terms,
$\a_{\b\cup\cA}^*\sle \a_\b^*\a_\cA^*\al_\b^*$.
\end{lemma}

\begin{proof}
The proof is in three steps.

We first show (1) $\a_\cA^*\tgt_\b\sle \tgt_\b\a_\cA^*\tlt_\b^*$, by
induction on the number of $\cA$-steps. Assume that $s\a_\cA^*
t'\a_\cA t\tgt_\b u$. By Lemma~\ref{prop-post}, there are $v$ and $v'$
such that $t'\tgt_\b v\a_\cA^* v'\tlt_\b u$. By induction hypothesis,
there are $s'$ and $s''$ such that $s\tgt_\b s'\a_\cA^* s''\tlt_\b^*
v$. Then, by Lemma~\ref{lem-commut}, there is $t''$ such that
$s''\a_\cA^* t''\tlt_\b^* v'$. Thus, $s\tgt_\b s'\a_\cA^* t''\tlt_\b^*
u$.

We now show (2) $\a_\cA^*\tgt_\b^*\sle
\tgt_\b^*\a_\cA^*\tlt_\b^*$, by induction on the number of
$\tgt_\b$-steps. Assume that $s\a_\cA^* t\tgt_\b u'\tgt_\b^* u$. After
(1), there are $s'$ and $t'$ such that $s\tgt_\b s'\a_\cA^*
t'\tlt_\b^* u'$. By strong confluence of $\tgt_\b$, there is $v$ such
that $t'\tgt_\b^* v\tlt_\b^* u$, where $t'\tgt_\b^* v$ is no longer
than $u'\tgt_\b^* u$. Hence, by induction hypothesis, there are $s''$
and $t''$ such that $s'\tgt_\b^* s''\a_\cA^* t''\tlt_\b^*
v$. Therefore, $s\tgt_\b^* s''\a_\cA^* t''\tlt_\b^* u$.

We now prove (3) $(\tgt_\b\cup\a_\cA)^*\sle
\tgt_\b^*\a_\cA^*\tlt_\b^*$, by induction on the length of
$(\tgt_\b\cup\a_\cA)^*$. Assume that $s \a_{\tgt_\b\cup\cA} t
\a_{\tgt_\b\cup\cA}^* u$. There are two cases. First, $s\tgt_\b
t$. This case follows directly from the induction hypothesis. Second,
$s\a_\cA t$. By induction hypothesis, there are $t'$ and $u'$ such
that $t\tgt_\b^* t'\a_\cA^* u'\tlt_\b^* u$. After (2), there are $s''$
and $u''$ such that $s\tgt_\b^* s''\a_\cA^* u''\tlt_\b^* u'$. Finally,
by Lemma~\ref{lem-commut}, there is $t''$ such that $u''\a_\cA^* t''$
and $t'\tlt_\b^*$. Hence, $s\tgt_\b^* s''\a_\cA^* t''\tlt_\b^* t$.

We conclude by the fact that $\tgt_\b^* = \a_\b^*$.
\end{proof}

We now turn to the proof of 
${\a_{\b\cup\cB}^*} \sle {\a_\b^*\a_\cA^*\al_\b^*}$ 
on {\em $\cR$-stables} sets.

\begin{lemma}
\label{lem-BtoA-proof}
Let $\cR$ be an arity-compliant semi-closed left-linear algebraic
system. On any set of $\cR$-stable terms, $\a_{\b\cup\cB}^*\sle
\a_\b^*\a_\cA^*\al_\b^*$.
\end{lemma}

\begin{proof}
We first prove (1) $\a_{\cB_1} \sle \a_\b^*\a_\cA^*\al_\b^* $.
Let $R$ be the binary relation such that for all $s,t \in \cT$,
\[
R(s,t) ~\Alr~ [~ {s \a_{\cB_1}t} \A {s \a_\b^*\a_\cA^*\al_\b^* t} ~] ~.
\]
We have to show that $R$ is compatible with term-formation rules
and that for all $\vd = \vc \sgt l \a r \in \cR$, 
for all substitution $\s$, if $\vd\s \ad_{\cA \cup \b} \vc\s$
then $R(l\s,r\s)$ holds.
We only show this latter property.
Let $\vd=\vc\sgt l\a r$ be
a rule and assume that $l\s\a_{\cB_1} r\s$. Then,
$\vd\s\ad_{\b\cup\cA}\vc\s$. Since $\vc$ is a closed algebraic term,
we have $\vd\s\a_{\b\cup\cA}^*\vu\al_\cA^*\vc$ with both $\vc$ and
$\vu$ in $\b$-normal form. 
Because $\vd\s$ is stable, we can apply Lemma~\ref{lem-post-star-proof} and
obtain $\vv$ such that $\vd\s\a_\b^*\vv\a_\cA^*\vu$. By Prop.
\ref{prop-beta-alg}, there is $\s'$ such that $\vv\a_\b^* d\s'$. Now,
by Lemma~\ref{lem-commut}, $\vd\s'\a_\cA^*\vu$. Therefore, $s\a_\cA
t$.

We now prove (2) $\a_{\b\cup\cB_1}^*\sle \a_\b^*\a_\cA^*\al_\b^*$, by
induction on the number of $\b\cup\cB_1$-steps. Assume that
$s\a_{\b\cup\cB_1}^* t\a_{\b\cup\cB_1} u$. By induction hypothesis,
$s\a_\b^* s'\a_\cA^* t'\al_\b^* t$. There are two cases. First,
$t\a_\b u$. By $\b$-confluence, $t'\a_\b^* u'\al_\b^* u$. 
Applying Lem.~\ref{lem-post-star-proof}
leads to $s'\a_\b^*\a_\cA^*\al_\b^* u'$ and we get
$s\a_\b^*\a_\cA^*\al_\b^* u$. Assume now that $t\a_{\cB_1} u$. From
(1) it follows that,
$t\a_\b^* t'_2\a_\cA^* u'\al_\b^* u$. Then, by virtue of
$\b$-confluence, $t'\a_\b^* 
t''\al_\b^* t'_2$. Commutation of $\b$ and $\cA$ (Lem.~\ref{lem-commut})
gives $u''$ such that $t''\a_\cA^* u''\al_\b^*
u'$. Finally, by Lemma~\ref{lem-post-star-proof}, $s'\a_\b^*\a_\cA^*\al_\b^*
u''$. Therefore, $s\a_\b^*\a_\cA^*\al_\b^* u$.

We then prove by induction on $i\ge 1$ that 
$\a_{\cB_i}\sle \a_{\cB_1}$. Let $i \geq 1$
and let $P_i$ be the binary relation such that for all $s,t \in \cT$, 
\[
P_i(s,t) ~\Alr~ [~ {s \a_{\cB_i} t} \A {s \a_{\cB_1} t}~] ~.
\]
We have to show that $P_i$ is compatible with term-formation rules
and that for all $\vd = \vc \sgt l \a r \in \cR$, 
for all substitution $\s$, if $\vd\s \ad_{\cA \cup \b} \vc\s$
then $P_i(l\s,r\s)$ holds.
We only show this latter property.
Let $\vd=\vc\sgt l\a r$ be a rule and assume that
$l\s \a_{\cB_i} r\s$. Then,
$\vd\s \ad_{\b\cup\cB_{i-1}} \vc\s$. By induction hypothesis and since
$\vc$ is a closed algebraic term, we have
$\vd\s\a_{\b\cup\cB_1}^*\vu\al_\cA^*\vc$ with both $\vc$ and $\vu$ in
$\b$-normal form. 
By (2), $\vd\s \a_\b^*\a_\cA^* \vu \al_\cA \vc$. Therefore, $l\s \a_{\cB_1} r\s$.
\end{proof}

\begin{theorem}
Assume that $\cR$ is an arity-compliant semi-closed left-linear algebraic
system. If $\a_\cA$ is confluent then $\a_{\b\cup\cB}$ is confluent on any
set of $\cR$-stable terms.
\end{theorem}

\begin{proof}
Let $S$ be a stable set of terms and let $s\in S$ such that
$u\al_{\b\cup\cB}^* s\a_{\b\cup\cB}^* t$. By lemma~\ref{lem-BtoA-proof},
there are $u',s_1',s_2'$ and $t'$ such that $u\a_\b^* u'\al_\cA^*
s_1'\al_\b^* s$ and $s\a_\b^* s_2'\a_\cA^* t'\al_\b^* t$. In other
words, $u'\alr_{\b\cup\cA}^* t'$. Since $\cA$ is confluent, by Theorem
\ref{thm-conflA}, there is $s''$ such that $u\a_\b^* u'\a_{\b\cup\cA}^*
s''\al_{\b\cup\cA}^* t'\al_\b^* t$. We conclude by the fact that
$\a_\cA\sle\a_\cB$.
\end{proof}

%%%%%%%%%%%%%%%%%%%%%%%%%%%%%%%%%%%%%%%%%%%%%%%%%%%%%%%%%%%%%%%%%%%%%%%%%%%
% Proofs of section orthonormal
%%%%%%%%%%%%%%%%%%%%%%%%%%%%%%%%%%%%%%%%%%%%%%%%%%%%%%%%%%%%%%%%%%%%%%%%%%%

\section{Proofs of Section \ref{sec-exclus}}

We begin by the commutation of $\a^*_\b$ and $\a^*_{\cB_i}$.

\begin{lemma}
\label{lem-com-reg}
If $\cR$ is an orthonormal system and $\a_{\b\cup\cB_i}$ is confluent
then $\a^*_{\cB_{i+1}}$ and $\a^*_\b$ commute.
\end{lemma}

\begin{proof}
The proof follows the lines of the proof of Lemma~\ref{lem-commut}.
It also uses the relation $\tgt_\b$ defined in Sect.~\ref{sec-Aconfl}.
We just prove that if $\vd=\vc\sgt l\a r$ is a
rule such that 
$u\tlt_\b l\s\a_{\cB_{i+1}} r\s$ then there is a $v$ such that
$u\a_{\cB_{i+1}}^* v\tlt_\b r\s$. As $l$ is a non variable linear
algebraic term, there is a substitution $\s'$ such that $\s\tgt_\b
\s'$ and $l\s\tgt_\b l\s'=u$. Therefore, $r\s\tgt_\b r\s'$.  It
remains to show that $l\s'\a_{\cB_{i+1}} r\s'$. Recall that
$\vd\s\a_{\b\cup\cB_i}^* \vc$. As $\vd\s\a_{\b}^* \vd\s'$, by
hypothesis, there is $\vv$ such that $\vd\s'\a_{\b\cup\cB_i}^*
\vv\al_\b^* \vc$.  But $\vc$ are $\b$-normal forms, hence $\vv=\vc$.
Therefore, $l\s'\a_{\cB_{i+1}} r\s'\tlt_\b r\s$.
\end{proof}

%%%%%%%%%%%%%%%%%%%%%%%%%%%%%%%%%%%%%%%%%%%%%%%%%%%%%%%%%%%%%%%%%%%%%%%%%%%
% parallel moves
%%%%%%%%%%%%%%%%%%%%%%%%%%%%%%%%%%%%%%%%%%%%%%%%%%%%%%%%%%%%%%%%%%%%%%%%%%%

We now turn to parallel moves. Lemma~\ref{lem-par-moves}
is decomposed into Lemmas~\ref{lem:par:moves:beta:cond}
and~\ref{lem:scr:parallel:beta:cond}.
We denote by $\a^=$ the reflexive closure of a rewrite relation
$\a$.
  
\begin{lemma}
\label{lem:par:moves:beta:cond}
Let $\cR$ be an orthonormal system and $i,j\geq 0$. Assume that,
for all $n$, $m$ such that $\{n,m\} <_{mul} \{i,j\}$ diagram $(i)$
commutes. Let $\vd=\vc\sgt l\a r$ be a conditional rewrite rule in
$\cR$. Then, diagram $(ii)$ commutes.
\[
\begin{array}{cc}
\xymatrix{
  s \ar@{->}[r]^{\b \cup \cB_n}_{*} \ar@{->}[d]_{\b \cup \cB_m}^{*}
  & t \ar@{..>}[d]^{\b \cup \cB_m}_{*} \\
  u \ar@{..>}[r]_{\b \cup \cB_n}^{*}
  & v
}
&
\xymatrix{
  l\s \ar@{->}[r]^{\cB_i} \ar@{->}[d]_{\tgt_{\cB_j}}
  & r\s \ar@{..>}[d]^{\tgt_{\cB_j}} \\
  u \ar@{..>}[r]_{\cB_i}^{=}
  & v
} \\ (i) & (ii) \\
\end{array}
\]
\end{lemma}

\begin{proof}
  The results holds if $i=0$ since $\a_{\cB_0}=\emptyset$.
  If $j = 0$, then $u = l\s$ and take $v=r\s$.

  Assume that $i,j >0$ and
  write $q_1,\dots,q_n$ for the (disjoint) occurrences in
  $l\s$ of the redexes 
  contracted in $l\s \tgt_{\cB_j} u$.
  Therefore, for all $k$, $1 \leq k \leq n$, there exists a rule $\rho_k:
  \vd_k = \vc_k \sgt l_k \a r_k$ and a substitution $\t_k$ such that
  $l\s|_{q_k} =l_k\t_k$.
  Thus,  $ u= l\s[r_1\t_1]_{q_1}\dots[r_n\t_n]_{q_n}$.
  It is possible to rename variables and assume that $\rho$, $\rho_1, \dots
  \rho_n$ have disjoint variables. 
  Therefore, we can take $\s \equiv \t_1 \equiv \cdots \equiv \t_n$.
  
  Assume that there is a non-variable superposition, \ie that a $q_k$
  is a non variable occurrence in $l$.
  Hence rules $\rho$ and $\rho_k$ forms an instance of a critical pair
  $ \vd'\mu = \vc'\sgt ({l[r_k]_{q_k}}\mu,r\mu) $ and
  there exists a substitution $\mu'$ such that $\s = \mu
  \mu'$.
  By definition of orthonormal systems, $|\vd'\mu| \geq 2$ and there is
  $m\neq p$ such that $c'_m \neq c'_p$ and $d'_m\mu = d'_p\mu$.
  Let us write $h$ for $max(i,j) - 1$.
  As $d'_m\mu = d'_p\mu$ we have $d'_m\s = d'_p\s$ and it follows
  that $c'_m \al_{\b \cup 
  \cB_h}^* d'_m\s = d'_p\s \a_{\b \cup \cB_h}^* c'_p$.
  But $\{h,h \} <_{mul} \{i,j \}$ and by assumption $\a_{\b \cup \cB_h}$ is
  confluent. Therefore we must have $c'_m \ad_{\b \cup \cB_h} c'_p $.
  But it is not possible since $c'_m$ and $c'_p$ are distinct normal forms.
  Hence, conditions of $\rho$ and $\rho_k$ can not be both satisfied
  by $\s$ and 
  $\a_{\b \cup \cB_{h}}$ and it follows that there is no non-variable superposition.

  Therefore, each $q_k$ is of the form
  $u_k.v_k$ where 
  $l|_{u_k}$ is a variable $x_k$.
  Let $\s'$ be such that $\s'(x_k) = \s(x_k)[r_k\s]_{v_k}$ 
  and $\s'(y) = \s(y)$ if $y \not\equiv x_k$ for all $1 \leq k \leq n$.
  Then, $l\s \tgt_{\cB_j} l\s'$ and by linearity of $l$, $u=l\s'$.
  Furthermore, $r\s \tgt_{\cB_j} r\s'$.
  We now show that $l\s' \a_{\cB_i} r\s'$.
  We have $\vd\s  \a_{\b \cup \cB_{i-1}}^* \vc$ and $\vd\s
  \a_{\cB_j}^* \vd\s'$. 
  As $i,j > 0$, we have $\{i-1,j\} <_{mul} \{i,j\}$.
  Therefore, by assumption $\a^*_{\b \cup \cB_{i-1}}$ and 
	$\a^*_{\b \cup \cB_{j}}$
  commute and there exist 
  terms $\vc'$ such that $\vd\s' \a_{\b \cup
  \cB_{i-1}}^*  \vc'  \al_{\b \cup \cB_{j-1}}^*  \vc$.
  But as terms in $\vc$ are $\a_{\b\cup \cB}$-normal forms, we have $\vc'
  = \vc$ and it follows that $l\s' \a_{\cB_i} r\s'$.
\end{proof}

%%%%%%%%%%%%%%%%%%%%%%%%%%%%%%%%%%%%%%%%%%%%%%%%%%%%%%%%%%%%%%%%%%%%%%%%%%%
%
%%%%%%%%%%%%%%%%%%%%%%%%%%%%%%%%%%%%%%%%%%%%%%%%%%%%%%%%%%%%%%%%%%%%%%%%%%%

\begin{lemma}
\label{lem:scr:parallel:beta:cond}
Let $\cR$ be an orthonormal system and $i,j\geq 0$. Diagram $(iii)$
commutes if and only if for all rule $\vd=\vc\sgt l\a r$, diagrams
$(iv)$ and $(v)$ commute.
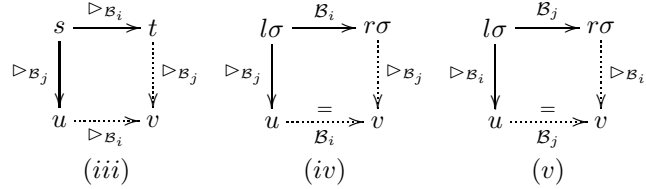
\[
\begin{array}{ccc}
  % (i)
  \xymatrix{
    s \ar@{->}[r]^{\tgt_{\cB_i}} \ar@{->}[d]_{\tgt_{\cB_j}}
    & t \ar@{..>}[d]^{\tgt_{\cB_j}} \\
    u \ar@{..>}[r]_{\tgt_{\cB_i}}
    & v
  }
  &  % (ii)
  \xymatrix{
    l\s \ar@{->}[r]^{\cB_i} \ar@{->}[d]_{\tgt_{\cB_j}}
    & r\s \ar@{..>}[d]^{\tgt_{\cB_j}} \\
    u \ar@{..>}[r]_{\cB_i}^{=}
    & v
  }
  &  % (iii)
  \xymatrix{
    l\s \ar@{->}[r]^{\cB_j} \ar@{->}[d]_{\tgt_{\cB_i}}
    & r\s \ar@{..>}[d]^{\tgt_{\cB_i}} \\
    u \ar@{..>}[r]_{\cB_j}^{=}
    & v
  }
  \\
  (iii) & (iv) & (v)
\end{array}
\]
\end{lemma}

\begin{proof}
The ``only if'' statement is trivial.
For the ``if'' case, let $s,t,u$ be three terms such that
$ u  \tlt_{\cB_j}  s  \tgt_{\cB_i} t $.
If $s$ is $t$ (resp. $u$), then take $v = u$ (resp. $v = t$).
Otherwise, we reason by induction on the structure of $s$.
If there is a rooted reduction, we conclude by commutation of diagrams
$(iv)$ and $(v)$.
Now assume that both reductions are nested.
If $s$ is an abstraction, we conclude by induction hypothesis.
Otherwise $s$ is an application $s_1 s_2$, and by assumption
$u = u_1 u_2$ and $t = t_1 t_2$ with $u_k \tlt_{\cB_j} s_k \tgt_{\cB_i} t_k$.
In this case also we conclude by induction hypothesis. 
\end{proof}

%%%%%%%%%%%%%%%%%%%%%%%%%%%%%%%%%%%%%%%%%%%%%%%%%%%%%%%%%%%%%%%%%%%%%%%%%%%
% shallow confluence
%%%%%%%%%%%%%%%%%%%%%%%%%%%%%%%%%%%%%%%%%%%%%%%%%%%%%%%%%%%%%%%%%%%%%%%%%%%

We now turn to the main result with orthonormal systems.

\begin{theorem}
If $\cR$ is an orthonormal system then $\a_{\b\cup\cB}$ is shallow confluent.
\end{theorem}

\begin{proof}
By induction on $<_{mul}$, we show the commutation of $\a^*_{\b\cup\cB_i}$
and $\a^*_{\b\cup\cB_j}$ for all $i,j\geq 0$.
The least unordered pair $\{i,j\}$ with respect to $<_{mul}$ is
$\{0,0\}$. As $\a_{\b\cup\cB_0}=\a_\b$ by definition, this case holds by
confluence of $\b$.

Now, assume that $i>0$ and that the commutation of $\a^*_{\b\cup\cB_n}$ and
$\a^*_{\b\cup\cB_m}$ holds for all $n,m$ with $\{n,m\} <_{mul} \{i,0\}$.  As
$\{i-1,i-1\} <_{mul} \{i,0\}$, $\a_{\b\cup\cB_{i-1}}$ is confluent and the
commutation of $\a^*_{\b\cup\cB_i}$ and $\a^*_{\b\cup\cB_0}$ ($=\a^*_\b$) follows from
lemma \ref{lem-com-reg}.

The remaining case is when $i,j>0$.  Using the induction hypothesis,
from Lemma \ref{lem:par:moves:beta:cond} and
\ref{lem:scr:parallel:beta:cond}, we obtain commutation of
$\tgt_{\cB_i}$ and $\tgt_{\cB_j}$, which in turn implies commutation
of $\a^*_{\cB_i}$ and $\a^*_{\cB_j}$. Now, as $\{i-1,i-1\} <_{mul} \{i,j\}$, by
Lemma \ref{lem-com-reg}, $\a^*_\b$ and $\a^*_{\cB_i}$ commute. This way, we also
obtain the commutation of $\a^*_\b$ and $\a^*_{\cB_j}$. Then, the commutation of
$\a^*_{\b \cup \cB_i}$ and $\a^*_{\b\cup\cB_j}$ easily follows.
\end{proof}

\end{document}